\documentclass[preprint,12pt]{elsarticle}

\usepackage{amssymb}
\usepackage{siunitx}
\usepackage{hyperref}

\usepackage{lineno}

\journal{iScience}

\begin{document}

\begin{frontmatter}



\title{An ensemble approach to the structure-function problem in microbial communities}


\author
{Chandana Gopalakrishnappa,$^{1,\dagger}$ Karna Gowda,$^{2,3,\dagger}$ Kaumudi Prabhakara,$^{2,3,\dagger}$ Seppe Kuehn$^{2,3\ast}$\\\vspace{0.3in}
\footnotesize{$^{1}$Department of Physics, University of Illinois at Urbana-Champaign,}\\
\footnotesize{Urbana, IL 61801, USA}\\
\footnotesize{$^{2}$Department of Ecology and Evolution, University of Chicago,}\\
\footnotesize{Chicago, IL 60637, USA}\\
\footnotesize{$^{3}$Center for the Physics of Evolving Systems, University of Chicago,}\\
\footnotesize{Chicago, IL 60637, USA}\\
\footnotesize{$^\dagger$These authors contributed equally to this work.}\\
\footnotesize{$^\ast$Correspondence: seppe.kuehn@gmail.com}
}

\begin{abstract}
The metabolic activity of microbes has played an essential role in the evolution and persistence of life on Earth. Microbial metabolism plays a primary role in the flow of carbon, nitrogen and other essential resources through the biosphere on a global scale. Microbes perform these metabolic activities in the context of complex communities comprised of many species that interact in dynamic and spatially-structured environments. Molecular genetics has revealed many of the metabolic pathways microbes utilize to generate energy and biomass. However, most of this knowledge is derived from model organisms, so we have a limited view of role of the massive genomic diversity in the wild on metabolic phenotypes. Further, we are only beginning to get a glimpse of the principles governing how the metabolism of a community emerges from the collective action of its constituent members. As a result, one of the biggest challenges in the field is to understand how the metabolic activity of a community emerges from the genomic structure of the constituents. Here we propose an approach to this problem that rests on the quantitative analysis of metabolic activity in ensembles of microbial communities. We propose quantifying metabolic fluxes in diverse communities, either in the laboratory or the wild. We suggest that using sequencing data to quantify the genomic, taxonomic or transcriptional variation across an ensemble of communities can reveal low-dimensional descriptions of community structure that can explain or predict their emergent metabolic activity. We motivate this approach broadly, and situate it historically. We survey the types of communities for which this approach might be best suited and then review the analytical techniques available for quantifying metabolite dynamics in communities. Finally, we discuss what types of data analysis approaches might be lucrative for learning the structure-function mapping in communities from these data.
\end{abstract}

\begin{keyword}
microbial communities \sep metabolism \sep dimension reduction \sep machine learning \sep genomic structure \sep 
\end{keyword}

\end{frontmatter}
\section*{Introduction: the structure-function problem in microbial communities.}
\label{S:1}
The evolutionary history of the biosphere is inextricably linked to the metabolic activities of microbes. Since life arose on this planet, microbes have lived in consortia 
that saturated nearly every biochemical niche on the planet, driving global transformations in the chemical composition of the biosphere via metabolic processes from fermentation to photosynthesis to respiration~\cite{falkowski_global_2000,Canfield:2010ib,nelson_global_2016,sunagawa_structure_2015,zakem_redox-informed_2020}. As such, microbes and the communities in which they reside are the result of an ongoing eco-evolutionary process that couples the transformation of metabolites to the complex dynamics of interacting ecological systems across many spatial and temporal scales.

Given the importance of the metabolic activity of microbial communities we argue that a major goal for the field should be to predict, design, and control the metabolism of microbial communities in complex, natural, and engineered settings. Accomplishing this goal requires understanding how the structure of a community, in terms of the taxa present and its genomic composition, determines its metabolic activity in a given environmental context. The sequencing revolution has revealed the structure of microbial communities at the level of the taxa present, the genes they posses and the dynamics of gene expression.
This means that we now have a detailed and dynamic ``parts list'' for microbial communities in terms of taxa and genomic composition across a range of environments, from anaerobic digesters~\cite{Bocher:2015im,TOERIEN1969385,vanwonterghem_deterministic_2014} to the human gut~\cite{blanton_gut_2016,raman_sparse_2019}, soils~\cite{bahram_structure_2018}, and the ocean~\cite{sunagawa_structure_2015}. For some metabolic processes we can interpret gene content and taxa in terms of the specific metabolic processes that they are capable of. For example, we know the dominant taxa that perform processes such as nitrification~\cite{Bock2013} or polysaccharide degredation~\cite{sanchez-gorostiaga_high-order_2019}. Further, by annotating metagenomic data we can assign specific functional roles for many (but not all) of the genes present in a given community. As a result, we can measure the prevalence of enzymes that perform the reactions necessary for specific metabolic processes. Despite the remarkable scale and breadth of these sequencing data, we still do not have a predictive, quantitative framework for using these data to understand, predict, and design the metabolism of the communities in complex environments. 

In this perspective we explore what makes this problem both challenging and important. We propose a specific approach to begin to address this question, and examine what types of communities and associated metabolic processes might be amenable to this approach. We review the techniques that are relevant to implementing the approach with a focus on methods for quantifying metabolites.


\subsection*{The significance of finding a solution}

Microbial communities play an outsized role in driving fluxes of nutrients through the biosphere. Photosynthetic microbes are responsible for nearly half of the carbon fixation on the planet~\cite{falkowski_role_1994}. These phototrophs work in concert with heterotrophic bacteria that enable primary productivity in terrestrial, marine and fresh water environments~\cite{kirchman_processes_2012,madigan_brock_2018}. We are only beginning to glimpse the role of the collective in this nearly 100 gigaton annual carbon flux. Bacteria and archaea in anaerobic environments degrade complex carbon sources to methane, playing an important role in carbon recycling and climate change~\cite{madigan_brock_2018}. 

In the nitrogen cycle, microbes play a key role in nitrogen fixation (dinitrogen gas to ammonia), nitrification (ammonia to nitrate), and denitrification (nitrate to dinitrogen gas)~\cite{stein_nitrogen_2016}. These processes are key for wastewater treatment~\cite{cydzik-kwiatkowska_bacterial_2016} and human health~\cite{turnbaugh_human_2007}. A critical challenge is to form a quantitative and predictive understanding of how microbial communities drive these fluxes. To give a concrete example, the process of denitrification, performed by bacterial communities in soils, reduces nitrate to dinitrogen gas. An intermediate in the conversion of nitrate to dinitrogen is the potent greenhouse gas and ozone depleting compound nitrous oxide~\cite{tian_comprehensive_2020}. Denitrifying communities in some cases (especially in agricultural soils) can leak nitrous oxide, but in other cases fully convert nitrous oxide to harmless dinitrogen gas. The question then becomes: what controls the production of nitrous oxide from denitrifying communities in soils? Can we manipulate these microbial communities to limit nitrous oxide production? To address this we need to understand how the structure of the community and the environmental context determine the flux of metabolites through the system. 

Similarly, the essential importance of resident microbiota in host health is now clear~\cite{turnbaugh_human_2007}, but as yet it is unclear how to rationally manipulate these communities to benefit the host. There exists tantalizing evidence that this can be done, for example by altering metabolic phenotypes~\cite{turnbaugh_obesity-associated_2006} or treating persistent infections~\cite{lawley_targeted_2012}, but we lack general approaches for developing such strategies. Here we focus on environmental microbiomes, but we emphasize that the strategy proposed here could, and in a few cases has been~\cite{raman_sparse_2019}, applied to host associated communities. 

\subsection*{Defining structure and function}
Before moving forward we take a moment to define community structure and function. We define the structure of the community as the taxa present, as well as the genomic structure of each taxon, which may include everything from the detailed knowledge of the regulatory architecture of each gene, to the syntenic organization of the genome~\cite{junier_universal_2018}, to even the presence of phage. The structure of the community may, if necessary, include transcriptional or proteomic information at the metagenomic or single-taxon level as well. 

We define the function of a community as the collective metabolic activity of all constituent organisms, which therefore operates in the space of metabolites. The dynamic or steady-state flux of metabolites through the consortium define its metabolic function. Depending on the context, the most important metabolic fluxes may include electron donors (e.g., organic carbon), acceptors (e.g., oxygen, nitrate), secondary metabolites, biomass, overall catabolic activity, or byproducts. A note about usage: some readers may find the term ``function" teleological, implying some sort of purpose on microbial communities. We use the term function to mean the \textit{activity or action} of a community without any implication of purpose. Despite this potential confusion, we find that using the term function is a useful shorthand. In particular, we would like to invoke a certain symmetry between the ideas presented here and the problem of sequence, structure and function at the level of proteins. The structure-function problem for microbial communities is therefore to deduce the mapping from the space of genes, transcripts, proteins and taxonomic organization to metabolite fluxes, and to understand the environmental dependence and context in which this mapping is relevant. 




\subsection*{How should we approach the problem?} Understanding how the metabolic activity of a microbial community emerges from the taxa present and their metabolic capabilities is a problem of connecting heirarchical scales of biological organization, from genes to phenotypes and interactions in specific environmental contexts. What makes this challenging is the fact that processes at different scales feed back on one another. For example, compounds that mediate interactions between strains can do so by modulating gene expression~\cite{beliaev_inference_2014}. Similarly, phenotypic variation in one organism can modify interactions~\cite{mickalide_higher-order_2019}, and organisms modify their environment with widespread impacts on other members of the consortium~\cite{ratzke_ecological_2018}.

One way to proceed is via the reductionist mode that has motivated biology over the last century~\cite{woese2004}. In the context of communities, this would mean dissecting the mechanistic and physiological metabolic properties of each member of the community and understanding how metabolite dynamics emerge at the level of the collective. The challenge is the complexity of these systems, which makes a detailed, mechanistic understanding of the collective metabolism a massive undertaking. For synthetic communities comprising model organisms, where detailed information are available, this type of approach has had some success ~\cite{orth_what_2010,harcombe_metabolic_2014}. For comparatively simple communities of a few strains, such as bacteria cross-feeding amino acids~\cite{wintermute_emergent_2010} or ciliates consuming bacteria~\cite{mickalide_higher-order_2019}, detailed models have been constructed and validated. However, in environmental or host-associated contexts, where the number of strains present is enormous and many organisms present are poorly studied or challenging to work with in the laboratory, this approach faces huge challenges.

In this scenario what are we to do?  On the one hand, building detailed models as described above is ill-advised. Even when such detailed models can be built, the challenge of distilling simple principles from these models increases with their complexity (see Borges ``On the exactitude of science''~\cite{borges_collected_1998}). On the other hand, we know from many examples that a huge number of processes from antibiotic warfare~\cite{vetsigian_structure_2011} to competition~\cite{Friedman:2017dk}, mutualism~\cite{hom_niche_2014}, and stress responses~\cite{amarnath_stress-induced_2021} influence interactions and metabolism. So how can we justify not building models that include these details?

In the spirit of Philip Anderson's influential essay~\cite{anderson_more_1972}, it may be that understanding communities requires explaining entirely new, and potentially simpler properties that are emergent at the level of the collective. In this case, due to the scale and complexity of the community, new and distinct phenomena emerge from the individual parts, and discovering the organizing principles requires an approach that goes beyond dissecting the detailed metabolic phenotypes of each individual member of the community. To be clear, we are not advocating the idea that reductionist approaches are not useful. Their utility is clear from many examples~\cite{jacob_genetic_1961,zumft_cell_1997,alon_robustness_1999,basan_universal_2020,amarnath_stress-induced_2021}. Instead, we are asking whether making headway on the ``structure-function problem'' as we have defined it might require a complementary approach that focuses not on the detailed mechanisms and physiology of each organism, but on patterns that are evident at the level of the collective. Here we discuss such an approach.

\subsection*{Learning the right variables: the power of statistics across ensembles}

We find it useful to cast the structure-function question in terms of a prediction problem. In this framing, the goal is to predict metabolite dynamics or fluxes from community structure for a given environment or set of environmental conditions. The key question then becomes, what structural elements must we quantify to predict function? Equivalently, we could ask, how predictive of metabolic function is community gene content, taxonomy, transcriptional or proteomic data? 

One approach to this problem, which has found success in both physics and biology, is to look for statistical regularities across replicate systems and allow these patterns to naturally define variables that can be used to make predictions. In many cases, this approach can reveal the salient, emergent, features of a system and provide deep insight into their function.

Specifically, from proteins~\cite{halabi_protein_2009} to multicellular organisms~\cite{alba_global_2021}, examining statistical variation across many replicates of a system, or an ensemble, has proven powerful. For example, covariation across ensembles of homologous proteins have been used to reveal which amino acids are in contact in the folded structure~\cite{russ_natural-like_2005}, and co-evolving groups of residues that correlate with enzyme function~\cite{halabi_protein_2009}. Careful analysis of behavioral variation in large numbers of microbes~\cite{jordan_behavioral_2013} and flies~\cite{berman_mapping_2014} suggest that apparently very complex behaviors can in fact be described by a relatively small number of elementary behavioral features~\cite{berman_mapping_2014,katsov_dynamic_2017}. Statistical analysis of morphological variation across higher organisms suggests that morphologies adhere to constraints that some believe are associated with specific functional capabilities~\cite{raup_theoretical_1965,shoval_evolutionary_2012}. A recent study of variation in patterning in the fly wing shows that a single mode of variation describes the response of the wing developmental program to genetic and environmental perturbations~\cite{alba_global_2021}. For a recent piece discussing why low-dimensionality might be an inherent property of evolved systems see Eckmann and Tlusty~\cite{eckmann_dimensional_2021}.

The common feature of all of these examples is that by judiciously studying the variation across a carefully chosen ensemble of systems, one can often discover simple, relatively low-dimensional features that enable the prediction of the functional properties of the system. Here we advocate a similar approach to communities.

\subsection*{An ensemble approach to the structure-function problem}

\begin{figure}
\centering\includegraphics[width=1\textwidth]{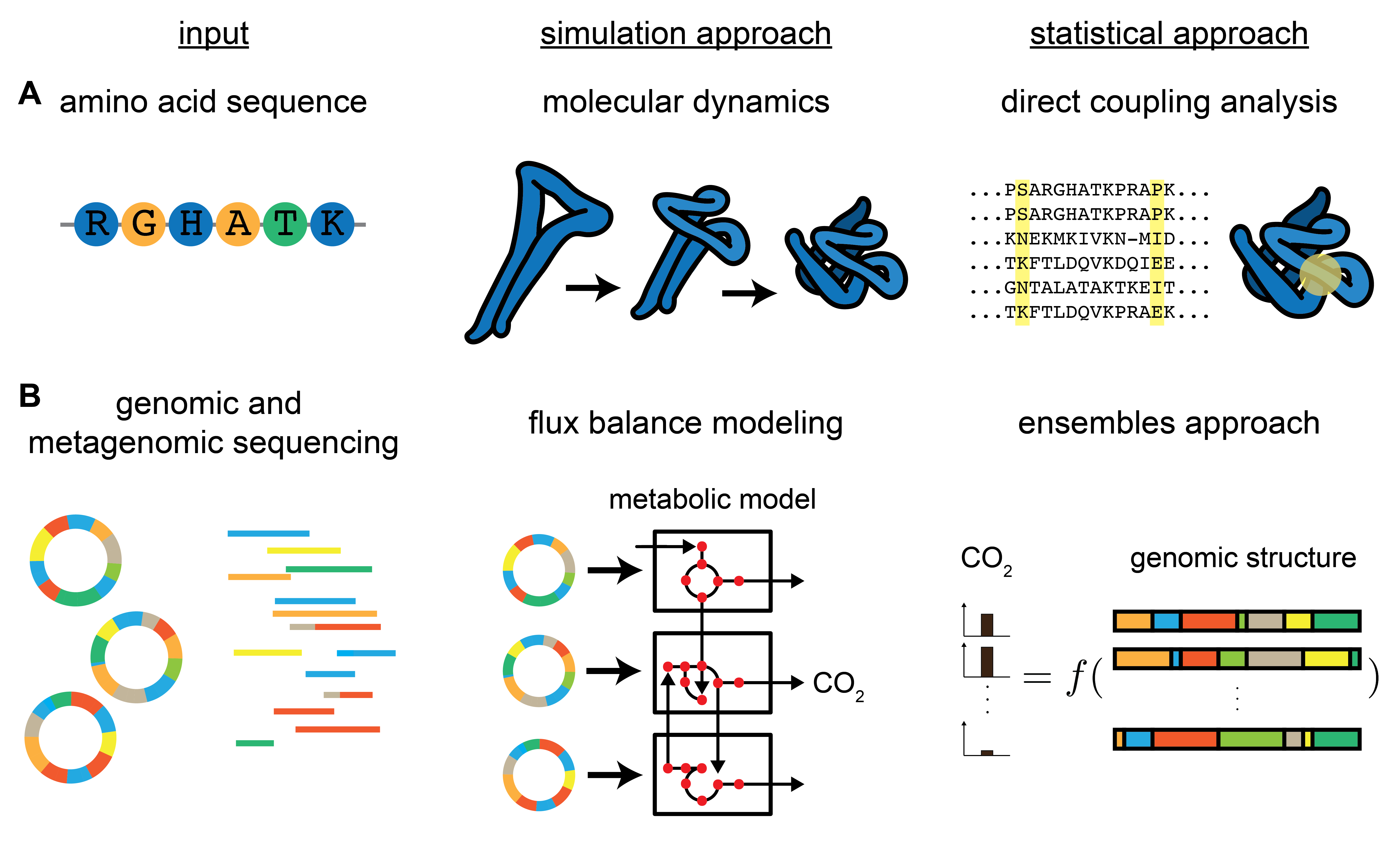}
\caption{\textbf{Sequence, structure and function in proteins and microbial communities.} We propose that there exist analogous solutions to the sequence-structure problem in protein folding and the structure-function problem in microbial communities. (A) The mapping from amino acid sequence to 3D protein structure can be accomplished either by a simulation approach (e.g., molecular dynamics) or by a statistical approach (e.g., direct coupling analysis). The former is a computationally-intensive strategy to simulate 3D protein structure based on first-principles modeling of atomic interactions. The latter leverages information about residue coevolution from an ensemble of amino acid sequences to infer which residues are in contact, allowing for an elegant and interpretable statistical inference of 3D structure. (B) The mapping from genomic and metagenomic sequences to community metabolic activity can be achieved through community flux balance modeling, or, as we propose, a statistical ensembles approach. The former requires genome-level metabolic models of each organism to be built, a labor-intensive iterative process that so far has been successful primarily in a handful of model organisms. The latter leverages the diversity and variation in an ensemble of communities to learn an an effective mapping between community sequence content metabolic activity.}
\label{Fig1}
\end{figure}

In light of these considerations, we propose an ensemble approach to the structure-function problem in communities. We motivate this approach by analogy to a similar approach taken at the level of proteins (Fig.~\ref{Fig1}). For proteins the structure function problem is to predict the fold and function of a protein from its amino acid sequence. One way to approach this problem is by performing detailed physico-chemical simulations of a polypeptide chain via molecular dynamics (Fig~\ref{Fig1}A). While much progress has been made in this approach, it has proven a huge technical challenge. However, a statistical approach, ignoring the mechanistic details and instead considering only statistics of a multiple sequence alignment does a remarkably good job of predicting protein folds~\cite{morcos_direct-coupling_2011}. Similar approaches reveal low-dimensional structure in proteins which is predictive of function~\cite{halabi_protein_2009,russ_evolution-based_2020}. These studies suggest that much can be learned by carefully considering variation across a suitably chosen ensemble of systems.

In the context of microbial communities flux balance models of community metabolism are analogous to molecular dynamics simulations in proteins because they attempt a detailed mechanistic accounting of all of the phenomena within a community (Fig.~\ref{Fig1}B). However, in communities there is comparatively little work pursuing a statistical approach analogous to the one taken at the level of proteins. 

Taking such an approach is precisely what we are advocating here. We propose quantifying the structure of a collection, or ensemble, of communities using sequencing while simultaneously measuring metabolite dynamics. Given such data, one can then approach the structure-function problem by asking whether variation in community structure (e.g., across metagenomes or metatransciptomes) permit quantitative insights into the functional properties of these communities. The proposal is then to leverage these insights to design, predict and control community function. Several studies in the past few years have begun to explore statistical approaches~\cite{Gowda:2021ve,raman_sparse_2019}. However, we suggest that this approach is under explored and that that there is a pressing need to collect new data sets that are explicitly designed to pursue a statistical approach to the structure-function problem in communities.



\subsection*{Overview of the paper}

We will focus on three main challenges that must be surmounted to apply the ensemble approach to the structure-function problem: (1) choosing an ensemble across which one should make comparisons and look for patterns, (2) measuring metabolite dynamics, which requires analytical chemistry techniques that are often not standard practice in microbial ecology labs, and (3) using these data to distill the mapping from structure to function.

Our intention is to provide a roadmap for how such an approach might be applied across communities of interest. We recognize that this roadmap is far from complete, and that many pitfalls exist that may render this approach challenging in various circumstances. 

We will not review sequencing technologies, which have been widely and capably recapped elsewhere~\cite{knight_best_2018}. We will focus largely on microbial communities in environmental contexts rather than health-related (human microbiome) contexts, in part because environmental microbiology is where our expertise lies, but also because of the abundance of existing literature on the latter topic. Finally, we will neglect the many recent advances in theoretical ecology, in particular the renaissance in consumer-resource models applied to communities, in service of focusing on questions that can be settled empirically.


\section*{Choosing communities and ensembles}
\label{S:2}

The first challenge is choosing a community and an associated metabolic process to interrogate. The choice is far from trivial and there is no simple prescription. Instead we appeal to the intuition of microbial ecologists, experts in physiology and the quantitative considerations of applied mathematics and physics. The goal should be to circumscribe a well-defined community and, if possible, an associated metabolic process where quantitative measurements in many replicates can be made. 

Choosing a metabolic process, and therefore specific metabolite measurements to be made, is a significant challenge, and compromises are inevitable. Here we reiterate the point that microbial communities are not ``functional'' in the sense that they are designed with some purpose. Despite this fact, as we will discuss below, the importance of metabolic activity of communities in specific niches is clear, as is the relevance of these processes for the biosphere more broadly. To make this point clear, we begin by discussing specific communities and their associated environments and metabolic processes, with an eye towards how one might apply the ensemble approach to dissecting their function.

\section*{Model microbial communities}
\label{S:3}
\subsection*{Structure-function in the wild}

\begin{figure}
\centering\includegraphics[scale=0.25]{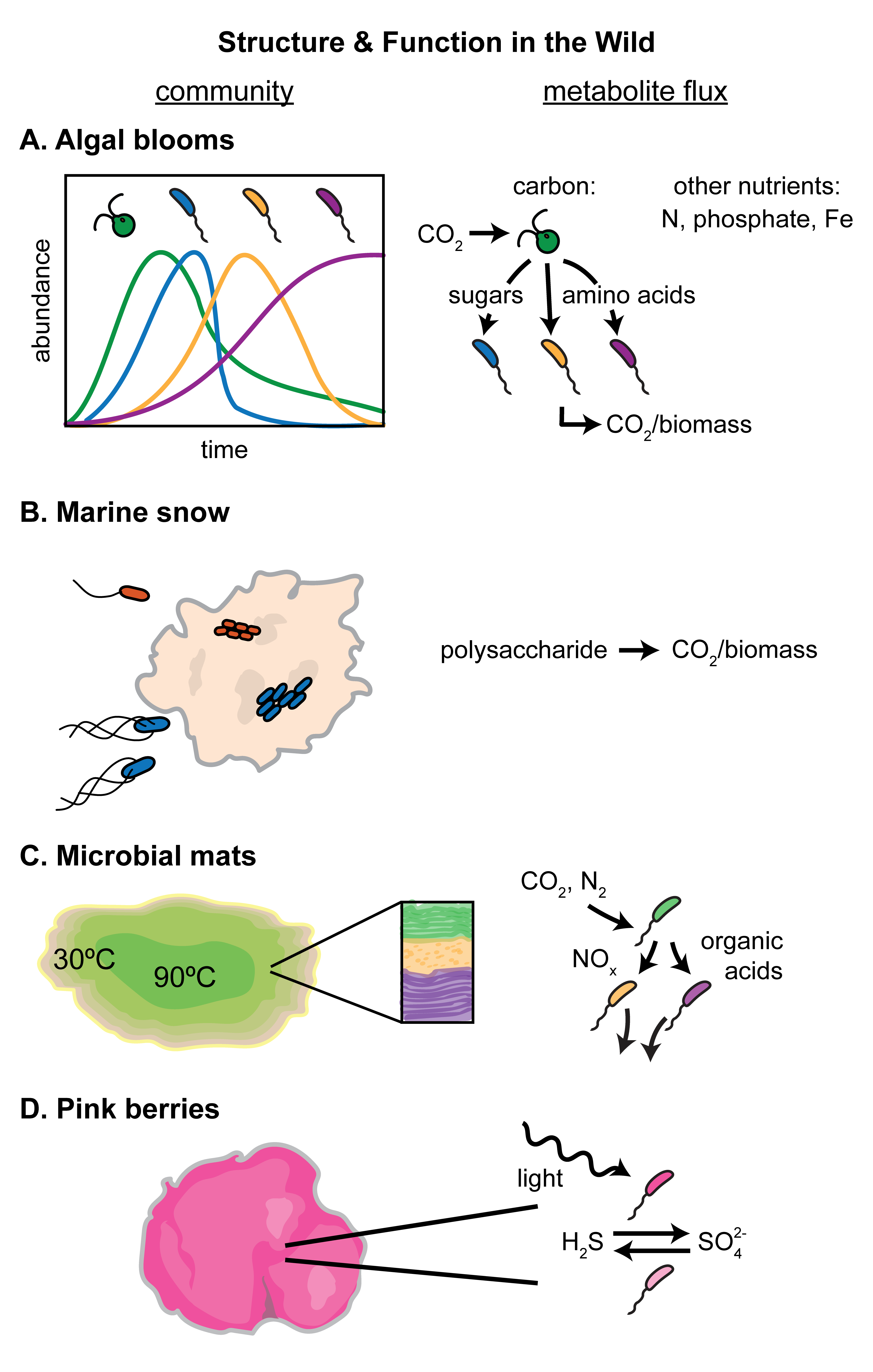}
\caption{\textbf{Community structure and function in the wild.} (A) Algal blooms are a microbial successional processes that follow from the input of exogenous nutrients to aquatic environments. Reduced carbon fixed from CO$_2$ by algae are consumed, along with other nutrients, by heterotrohic bacteria in reproducible successional dynamics. (B) Marine snow particles are aggregates of organic carbon that are formed near the ocean's surface and subsequently sink to the ocean floor. Microbial communities can degrade these particles, and the amount of carbon that is mineralized to CO$_2$ versus the amount that is sequestered on the ocean floor depends strongly on the structure of the microbial community. (C) Microbial mats are layered communities that occur at air-water interfaces, often in extreme thermal environments such as hot springs. The spatial structure of these communities follows from exchanges of nutrients governed by redox gradients. (D) Pink berries are microbial aggregates that cryptically (internally) cycle sulfur between photosynthetic purple sulfur bacteria and anaerobic sulfate reducing bacteria.} \label{fig:inthewild}
\end{figure}

\subsubsection*{Soils}
Perhaps no microbial community on Earth is more important than that which inhabits soils. The soil microbiome plays a key role in plant growth and physiology ~\cite{saleem_more_2019}, in particular through nitrogen fixing bacteria that provide reduced nitrogen for plant hosts. The storage of carbon, and production of CO$_2$ via respiration of reduced organic compounds in soils are key components of the global carbon cycle~\cite{lal_soil_2004,lal_soil_2004}. As the climate warms, so the rate of microbial respiration of CO$_2$ from soils increases~\cite{kirschbaum_temperature_1995,allison_soil-carbon_2010}, potentially driving a positive feedback loop with dire consequences for the global climate. Similarly, denitrification in agricultural soils is responsible for roughly \SI{80}{\percent} of the anthropogenic release of nitrous oxide (N$_2$O) (see \url{https://www.epa.gov/ghgemissions/overview-greenhouse-gases} and~\cite{tian_comprehensive_2020}). Nitrous oxide is 300-fold more potent than CO$_2$ as a greenhouse gas and responsible for approximately \SI{10}{\percent} of the global warming potential from human activity. For these reasons, there is keen interest in associating soil microbiome structure to process rates such as CO$_2$ or N$_2$O production and N$_2$ fixation. However, most attempts to find a relationship between soil microbiome structure and the rates of key metabolic processes in soils have found only marginal success~\cite{graham_microbes_2016,rillig_role_2019,rocca_relationships_2015,fierer_embracing_2017}.

Despite these difficulties we see reason for optimism. It is known that a relatively small number of environmental factors are the dominant drivers of variation in soil community structure: pH, moisture, carbon and nitrogen availability, temperature, and redox potential~\cite{fierer_embracing_2017}. Moreover, while soils are routinely cited as very complex microbial communities, much like the human gut they are typically dominated by a handful of taxa (e.g., \textit{Acidobacteria}, \textit{Verrucomicrobia}~\cite{fierer_embracing_2017,crits-christoph_novel_2018}), with most other strains present in relatively low abundances. Moreover, there are clear patterns in the abundances of bacteria and fungi in soils, with high biomass turnover environments like grasslands dominated by bacteria and low biomass turnover forests dominated by fungi~\cite{fierer_embracing_2017}.

As acknowledged in recent meta-studies~\cite{graham_microbes_2016}, one challenge in associating soil community structure to metabolic function is a lack of high quality data sets where process rates (e.g., CO$_2$ production) are measured in a large ensemble of soil communities. Exceptions to this include a recent survey of global topsoil microbiomes~\cite{bahram_structure_2018}, and microcosm studies documenting the role of multiple environmental perturbations applied to soils~\cite{rillig_role_2019}. Despite these advances, the consensus remains that predicting metabolic processes in soils from microbial community structure is challenging~\cite{fierer_embracing_2017}. However, we note that in some cases this conclusion is derived from meta-studies that aggregate data from different experiments or labs. Such comparisons can be challenging given systematic errors in sequencing measurements between protocols~\cite{mclaren_consistent_2019}, and the strong dependence of measurements like soil pH on the technique employed~\cite{miller_comparison_2010}.

Given these considerations we propose that one route forward is the judicious collection of data from large ensembles of soil communities followed by careful quantification of process rates and community structure in a consistent manner. We acknowledge that even in the presence of such data, relating community composition at the taxonomic, genomic or transcriptomic levels to metabolite fluxes may remain a challenge. It may be that soil taxa are not ``good variables" for understanding function, and that chemical or abiotic properties such as the redox state of available organic carbon are necessary~\cite{keiluweit_anaerobic_2017}. Regardless, the importance of understanding the metabolic activities of soil communities cannot be overstated.

\subsubsection*{Algal blooms}
In aquatic systems, exogenous inputs of nitrogen and phosphorous, often driven by human activity, result in the dramatic successional process of the algal bloom~\cite{Teeling2012} (Fig.~\ref{fig:inthewild}A), wherein photosynthetic microbes explode in abundance. The rise of phototrophic microbes brings with it a complex community of non-photosynthetic (heterotrophic) bacteria that form tight symbioses with the phototrophs. These phototroph-heterotroph communities cycle large quantities of carbon, with phototrophs fixing CO$_2$ to reduced organic carbon, which is in turn consumed by the associated heterotrophic bacteria. 

In this system, a natural framing of the structure-function problem would be to ask how the taxonomic composition of the community impacts the fixation of carbon and eventual bacterial biomass production. Teeling \emph{et al.}~\cite{Teeling2012} have found that specific classes of bacteria grow at different phases of the bloom, e.g., Alphaproteobacteria dominated the pre-bloom, as blooming commenced, Bacteroidetes increased more rapidly than others, and then Gammaproteobacteria grew much later. Concurrent metagenomic studies showed that the abundances of enzymes capable of degrading carbohydrates and sulfatases required to degrade sulfated algal polysaccharides increased in abundance during the bacterial succession. The degradation of the larger polysaccharides leads to the production of shorter organic compounds, which was revealed by the expression of the relevant transporters. These results, and those of other studies~\cite{kimbrel_host_2019,louati_structural_2015,mcfeters_growth_1978,parulekar_characterization_2017,ramanan_phycosphere_2015}, indicate a link between taxonomy of bacteria associated with certain photosynthetic strains during blooming events. 

These results suggest that these associations are at least in part determined by the carbon catabolic activity of the bacteria and the fixed carbon that phototrophs excrete (see also \cite{Buchan2014}). Taking an ensemble approach to this problem could be accomplished by recapitulating bloom dynamics in a laboratory context~\cite{riemann_dynamics_2000}, where the identity of the phototroph and the composition of the bacterial community could be manipulated and the resulting total carbon flux quantified~\cite{astacio_closed_2020}. Such studies might shed insights on how to control blooms in natural settings.

\subsubsection*{Marine snow}
Marine snow is a term used to describe micrometer to millimeter-scale aggregates in the ocean, which are typically made of detritus containing carbon and nitrogen in the form of polysaccharides, microbes, and inorganic substances. These particles form at the upper levels of the ocean and sink to the ocean floor, transporting carbon in a ``pump'' that removes carbon from the atmosphere over geologic timescales~\cite{Alldredge1988,Gralka2020}. Communities of bacteria and other microbes consume these aggregates as they sink (Fig.~\ref{fig:inthewild}B). Understanding how the structure of marine snow communities impacts the degradation of carbon and the production of metabolic byproducts (e.g., CO$_2$) is a critical question for understanding global carbon fluxes.

Studying these particles \emph{in situ} is challenging~\cite{Kiorboe2007}. To circumvent this difficulty, some studies use synthetic particles to study community assembly and function. One study used agar beads \cite{Kiorboe2003} to study the colonization by microbes in raw seawater, revealing successional dynamics underlying particle degradation. The particles were initially colonized by bacteria, and later by flagellates. Cordero and colleagues have taken a similar approach to understand colonization dynamics~\cite{datta_microbial_2016,ebrahimi_cooperation_2019,Enke2019,pontrelli_hierarchical_2021}. These studies shed light on the metabolic roles played by different players during particle colonization dynamics: primary degraders excrete enzymes to break down polysaccharide chains, which creates a niche for secondary degraders capable of consuming monomers and oligomers of the polysaccharide, which in turn makes way for scavengers that take up metabolic byproducts of the primary and secondary degraders (ammonia, amino acids). 

In terms of the approach proposed here for mapping structure to function, particle degradation offers a powerful model system because so much is known about how the communities collectively degrade the particle. For example, it would be interesting to see whether one could take a statistical approach to inferring the key metabolic traits of primary and secondary degraders and scavengers from sequencing data alone. In this case the function of the community to predict would be the fraction of carbon degraded or the total CO$_2$ respired. The power of this system is the ability to manipulate structure, but the particles make it challenging to measure carbon degradation directly (see the next section for further discussion).

\subsubsection*{Plastisphere}
Closely related to the degradation of polysaccharide particles in the ocean is the recent rise of plastic debris in freshwater and marine environments, and the microbial communities associated with its degradation \cite{Amaral-Zettler2020}. Given the fact that a few million tons of plastic enter the ocean per year \cite{Jambeck2015}, this is an important ecosystem to study, not only to understand how these pollutants affect the ecosystem, but also to find potentially efficient plastic degrading communities. Further, plastic is a comparatively new environment on evolutionary timescales, making it interesting to study from the perspective of evolution. It has been shown that the community composition on plastic is different from the composition on other substances in the same conditions, and that certain taxa are commonly found on plastics \cite{Dussud2018,Kirstein2019}. Diatoms have been observed in high numbers, although a strong succession dynamics was observed. Other photoheterotrophs, heterotrophs, ciliates, fungi, and pathogens have also been observed. Although the functional role of these microbes are unclear, it is speculated that chemotaxis, interactions with metals and degradation of low molecular weight polymers are important factors that determine community composition \cite{Amaral-Zettler2020}. In this system the structure-function problem is similar to that outlined for marine snow - how does community structure determine the rate of plastic degradation?



\subsubsection*{Microbial mats}
Microbial mats are stratified communities that often form at air-water interfaces in extreme environments such as hot springs~\cite{Klatt2013}. Mat communities harbor a top layer of photoautotrophic bacterium (typically \textit{Synechococcus} sp.) that use light to fix carbon during the day and often fix nitrogen at night. Below the top layer are strata of various heterotrophs and anaerobic autotrophs~\cite{ward_natural_1998} (Fig.~\ref{fig:inthewild}C). These communities have been studied for decades, and much is known about the metabolic roles each strain plays in the community~\cite{bateson_photoexcretion_1988,anderson_formation_1987} and metagenomic datasets are available~\cite{lee_metagenomics_2018}. It is remarkable that similar mat structures form in many different contexts across the globe. In these mats much of the nutrients are fixed from CO$_2$ and N$_2$~\cite{steunou_situ_2006}, heterotrophic community members then consume reduced organic carbon excreted by the autotrophs. In this context the question of relating structure to function falls to mapping the flux of C, N and other metabolites through the mat to the taxonomic and metagenomic structure of the system. For example, what is the simplest community that will stably form a mat?  What pathways, for example in the primary autotroph, are essential to mat formation and which are dispensable? Further, given that mats support remarkable allelic diversity driven by extensive recombination~\cite{rosen_fine-scale_2015}, how is the functional genetic repertoire of these communities maintained? Preliminary work on mats in California \cite{Lee2018} showed the presence of a core genome across samples, and various other genes thought to be useful for specialized functions, but further metagenomic analysis is likely to be useful in addressing the structure-function question in mats. The main two challenges in these communities are obtaining axenic isolates from the mats and making high quality metabolite dynamics measurements \emph{in situ}. Petroff and colleagues have recently made exquisite quantitative measurements of oxygen dynamics in mats~\cite{petroff_subsurface_2017,tejera_oxygen_2018}, addressing the latter challenge, which opens the door to making quantitative links between community composition and metabolic activity.


\subsubsection*{Cryptic sulfur cycling in microbial aggregates}
Microbial metabolism drives the global cycling of sulfur through energy-generating oxidation and reduction reactions. It has become increasingly evident that much of this cycling occurs cryptically~\cite{Canfield:2010ib, Callbeck:2018iw} (i.e., with low steady state metabolites levels but substantial fluxes), often in the context of cellular aggregates where oxidation and reduction reactions occur in physical proximity~\cite{Wilbanks:2014kx, Callbeck:2018iw}. Inferring the presence and rate of cryptic sulfur cycling in microbial aggregates has important implications for our understanding of other elemental cycles, since sulfur cycling is often tightly coupled with the carbon and iron cycles. 

Remarkable examples of cryptic sulfur cycling phenomenon are the so-called ``pink berry’’ consortia~\cite{Seitz:ht,Wilbanks:2014kx}, which are discrete macroscopic ($\sim$1 cm) aggregates that occur on the surface of submerged sediments in the intertidal pools of Sippewissett Salt Marsh (Massachusetts, USA) (Fig.~\ref{fig:inthewild}D). The bright pink coloration of these aggregates is attributable to the purple sulfur bacteria (PSB) that make up the majority of cellular biomass. These PSB oxidize sulfide to sulfate via the process of anoxygenic photosynthesis. Accompanying these PSB are sulfate reducing bacteria (SRB), which derive energy from anaerobic respiration by catalyzing the reverse process, reducing sulfate to sulfide. Together these PSB and SRB are capable of locally and cryptically cycling sulfur via the syntrophic exchange of oxidized and reduced sulfur compounds~\cite{Wilbanks:2014kx}. 

While culture-based approaches to characterizing and reconstituting the pink berry consortia in the lab remain a challenge, the discrete nature of the pink berries presents an opportunity to statistically characterize the structure-function relationship at the level of individual aggregates harvested from the wild. Bulk metagenomic sequencing of pink berry consortia indicates that typically 2-3 phylotypes make up the majority of biomass in the aggregates~\cite{Wilbanks:2014kx}. Laboratory measurements of sulfide oxidation or sulfate reduction rates using spatially-explicit microprobe measurements followed by 16S metagenomic sequencing on harvested pink berries could provide insight into how the abundance of these phylotypes quantitatively relates to sulfur cycling. More generally, such an approach could be applied to characterizing cryptic sulfur cycling in other aggregated contexts, such as in marine particles~\cite{Callbeck:2018iw}. Predicting sulfur cycling is important to understanding other critical elemental cycles: sulfide oxidation by PSB can contribute substantially to carbon fixation~\cite{Dyksma:2016kh}, and sulfate reduction by SRB can be an important sink for reduced carbon~\cite{Liamleam:2007gc} and iron~\cite{Enning:2014ew} in environments where electron acceptors are otherwise scarce.

\subsection*{Community structure and function under domestication}

\begin{figure}
\centering\includegraphics[scale=0.4]{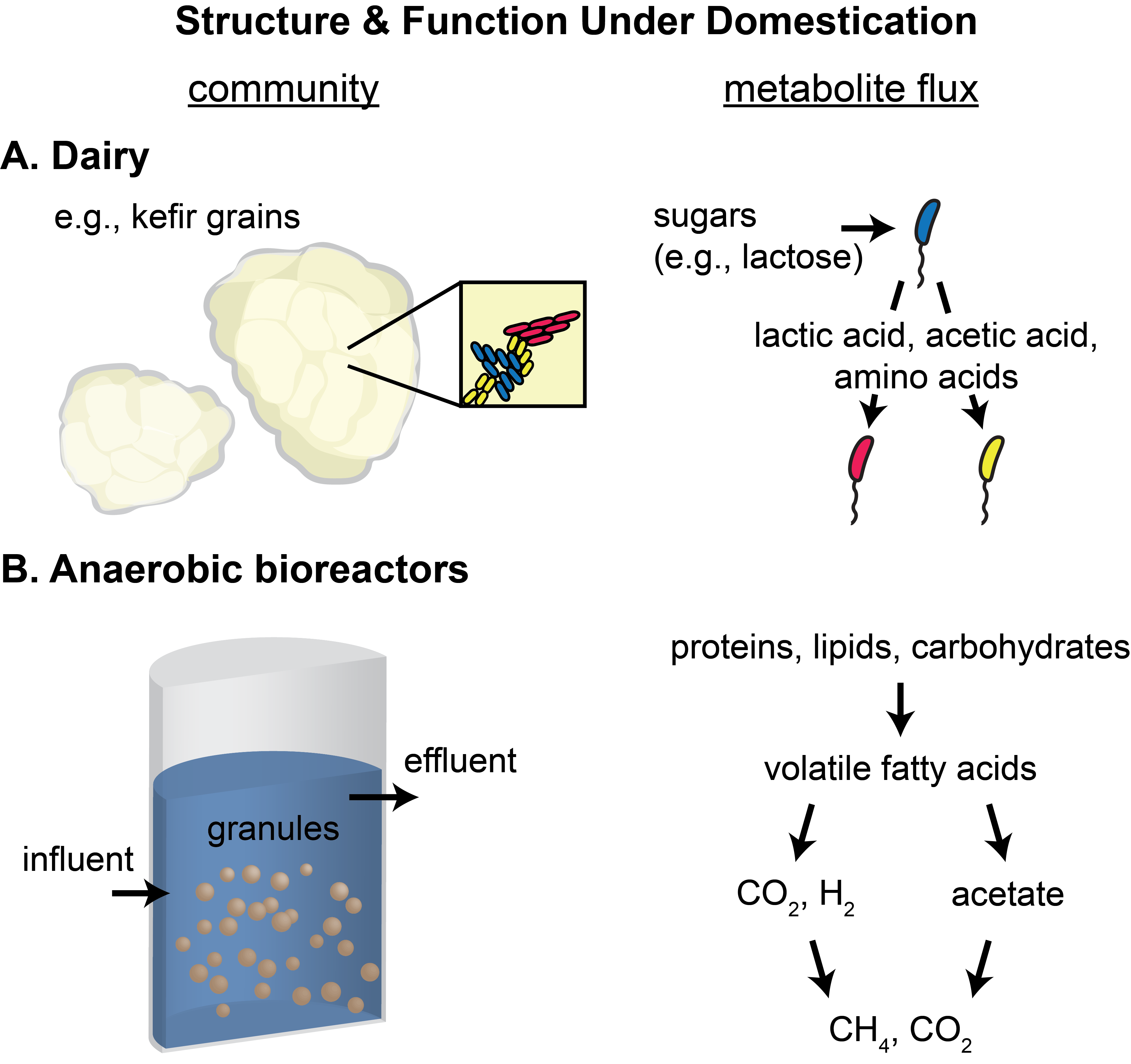}
\caption{\textbf{Community structure and function under domestication.} (A) Kefir grains extracellular-polymeric aggregates host to microbial communities that inoculate milk for the production of kefir. These communities undergo a reproducible successional process that involves the production and consumption of fermentation byproducts, which ultimately give kefir its desired flavor. (B) Anaerobic bioreactors often use granulated microbial communities to remove waste products such as reduced carbon, nitrogen, and phosphorous from water. Improving the performance and efficiency of these systems through the ensembles-informed design of communities would increase their viability as alternatives to traditional wastewater treatment approaches that expend significant energy on aeration.}\label{fig:domestication}
\end{figure}

Many key insights in the Origin of Species came from studying trait variation under domestication. In the same way, we propose that learning the rules for mapping structure to function in communities should leverage the many instances in which communities have been domesticated. Here we explore some of these opportunities.

\subsubsection*{Microbial communities in the dairy industry}
The production of yogurt, cheese and other dairy products relies heavily on microbes. The work of Dutton and colleagues on cheese rinds \cite{Wolfe2014} is an important example in this context. This study compared and characterized more than 100 cheese rinds from across 10 countries, and found that less than 15 bacterial taxa and 10 fungal taxa were present in abundances of more than $1\%$. Further, most of these taxa are not present in the starting cultures, and their function is unknown. They found that the community composition is strongly correlated to the aging process and moisture rather than geography or milk source. The functional profile of the communities, found using shotgun metagenomics, was correlated to pH as well, and pathways were shown to correlate as expected with the cheese types. These results point to the idea that the chemical environment is perhaps the strongest determinant of community structure.

Remarkably, when the dominant taxa were reconstituted \emph{in vitro} and cultured as a community in media to mimic treatments in different types of cheese rinds, divergent communities were formed depending on the treatments, which showed some properties similar to the original cheese rinds including their abundance dynamics. This remarkable result shows that these communities can be reconstituted in the laboratory to recapitulate some of the basic functional features of the domesticated cheese communities. It would be interesting to extend these studies further by looking quantitatively at the metabolite dynamics. The fact that these communities can be so readily manipulated means that learning the relationship between composition and metabolic activity is now accessible. What remains unclear, to us, is what the salient metabolic features of these communities are that should be explained. One way to approach this question would be to chemically characterize the transformations that occur in specific rinds and to then ask whether subsets of the full community can or cannot recapitulate these processes. 

Another promising microbiome in the dairy sector is the kefir community (Fig.~\ref{fig:domestication}A). Kefir grains are used to make a fermented drink like yogurt. They have about 50 bacterial and yeast taxa, which are resilient to stresses, and most of which perform lactic and acetic acid fermentation.  A recent study \cite{Blasche2021} found that kefir grains, which are polysaccharide matrices synthesized by the bacterial consortium, collected from diverse locations had a very similar core community and differed only in rare species. Kefir grains are sustained much like sourdough starters and added to milk to initiate a fermentation process. When added to milk the composition of the grain community is stable while the community present in the milk exhibits a succession. Metabolite changes during the colonization showed similar succession dynamics. The study dissects the interactions between the community on the grains and that in the milk, and demonstrates reproducible metabolite dynamics in this system. As a result, kefir constitutes another powerful platform for manipulating community structure (e.g., composition of the community on the grains) and learning the impact of those changes on the metabolite dynamics. Blasche \emph{et al.} have already made significant progress in this regard, but it remains to investigate in a high-throughput statistical fashion, how the composition of the grain community confers the remarkable robustness they observe.



\subsubsection*{Anaerobic bioreactors}
The treatment of wastewater for reuse and release into the environment requires the removal of large quantities of organic matter, much of which is insoluble or otherwise slow to degrade. Anaerobic bioreactors serve an important role in this industrial process, harnessing microbial metabolism to degrade such organic matter into CO$_2$ and CH$_4$ gases (Fig.~\ref{fig:domestication}B). Because the bioreactors are fed a range of inputs and are operated under a variety of conditions, the microbial communities that populate bioreactors are highly functionally and taxonomically diverse~\cite{Werner:2011hm}. Organisms that excrete extracellular enzymes degrade insoluble polymers into soluble monomers, while fermenters consume these compounds and excrete products including acetate and H$_2$. Methanogenic archaea can then ferment these products to produce CH$_4$~\cite{TOERIEN1969385}. Many functional attributes are used in practice to characterize the performance of anaerobic bioreactors, including removal of chemical oxygen demand (COD, a proxy for aerobically metabolizable matter in a water sample) and methanogenic activity. The resilience of a bioreactor community to input fluctuations has also been of interest~\cite{Fernandez:2000cq, Hashsham:2000fw, Werner:2011hm}.

Several studies have explored the statistical relationship between community taxonomic structure and methanogenic bioreactor performance. In an important early study, Tiedje and colleagues found that, under constant conditions, COD removal in a laboratory bioreactor was stable while community composition varied substantially over a two year period~\cite{Fernandez:1999ew}. This work suggested the role of functional redundancy~\cite{Louca:2018cv} in maintaining community function, and implied a degeneracy in the relationship between community taxonomy and function. However a more recent study of several industrial-scale bioreactors observed relative stability in both community taxonomy and reactor performance over a year-long period~\cite{Werner:2011hm}. Notably, variations in reactor performance were found to be related to community composition. This indicated that taxonomy is predictive of community function, though the authors argued that taxonomy is simply a proxy for functional genomic content due a close correspondence between phylogeny and metabolic function for organisms in anaerobic bioreactor systems. The conflict between these two results indicates that much is still unknown about the structure-function relationship in anaerobic bioreactors.

Recent work has advanced our understanding of structure and function in methanogenic bioreactors, leveraging a statistical ensembles approach to discover a predictive relationship between gene content and methanogenic activity~\cite{Bocher:2015im}. The authors generated 49 diverse enrichment cultures by seeding laboratory bioreactors with inocula taken from a large and eclectic collection of industrial-scale bioreactors. This ensemble enabled a linear regression approach to mapping methanogen activity (as measured by methane production rate per unit biomass) to variation in genomic content, specifically the abundance of sequence variants of a gene important to methanogenesis (\textit{mcrA}). Remarkably their approach produced a predictive model with relatively few variables, suggesting that only a few key \textit{mcrA} variants, or strains possessing these variants, are important for methanogenic activity. A powerful consequence of this approach is a prediction for which gene variants would improve the performance of an underperforming bioreactor. 

Given the importance and widespread use of bioreactors to process organic matter, these constitute important model systems. The field faces two important challenges. First, cultivating many of the slow growing taxa in these communities is difficult and this means that only equipped and experienced labs can readily work with these organisms. Second, the complexity of these communities makes carefully controlled and reproducible experiments a challenge, and as a result comparisons from one study to the next can be difficult. Standardizing conditions and starting inocula would therefore be a major advance.

\begin{figure}
\centering\includegraphics[scale=0.2]{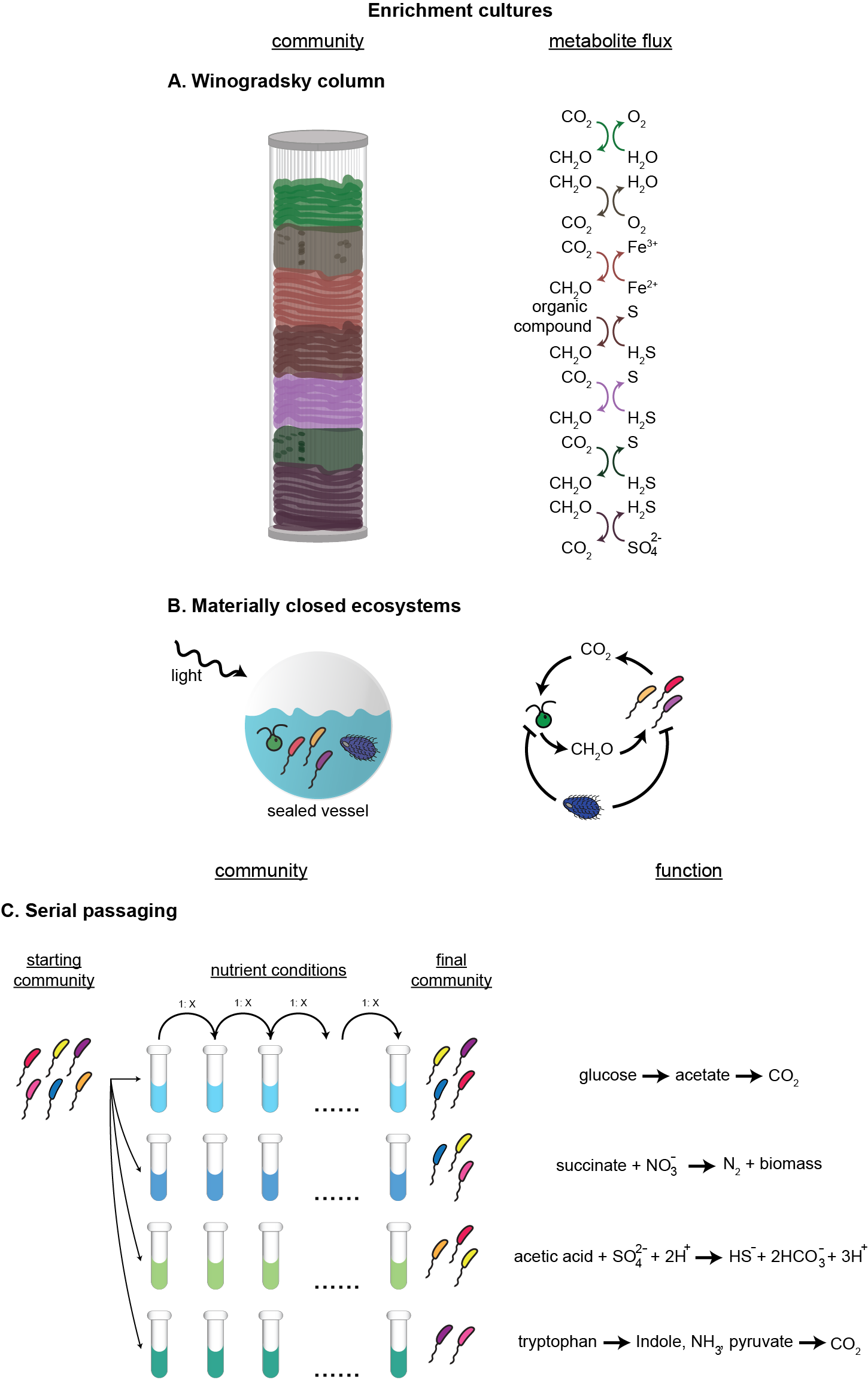}
\caption{\textbf{Bringing wild communities into the lab.} (A) Winogradsky columns are laboratory-assembled are lab-assembled communities with distinctive and reproducible spatially-stratified metabolite fluxes. These fluxes, and consequently community structure, arise from emergent redox gradients. (B) Materially closed ecosystems are communities grown in sealed vessels whose only energy input is light. Nutrient cycling in closed ecosystems arises from phototrophic organisms generating reduced carbon by fixing CO$_2$, which can then be  consumed by heterotrophic organisms. Predators such as ciliates can consume whole cells, facilitating the recycling of macromolecular biomass. (C) Serially-passaged communities enrich a complex environmentally-derived community on laboratory-controlled nutrient conditions (e.g., a fixed carbon source). The resulting communities are typically low-complexity, and demonstrate reproducible trophic roles and patterns of nutrient exchange.}
\label{Fig4}
\end{figure}

\subsection*{Bringing wild communities into the lab: enrichment cultures}
Another route to studying communities that is similar to domestication is to bring complex communities from nature into the laboratory and culture them in defined conditions. While these experiments allow the experimentalist to control the growth and incubation conditions, they often result in a drastic loss of diversity. As a result, this approach is likely poorly suited to understanding the structure of natural systems, but none the less should enable important insights into engineering and controlling communities. 

\subsubsection*{Winogradsky columns}
One of the most well-known methods for enrichment culture is a Winogradsky column (Fig.~\ref{Fig4}A). The method was developed by Sergei Nikolaievich Winogradsky to study (and discover) chemosynthesis, a process where energy is derived from inorganic compounds \cite{Zavarzin2006}. Winogradsky columns are glass cylinders (or bottles) loaded with a sediment, supplemented an organic carbon source (typically paper) and a sulfur source, that is then sealed off and illuminated. With time, different microbes occupy different levels of the column based on their metabolic capabilities. Each layer is characterized by distinct redox reactions (electron donor/acceptor pairs) that support metabolism locally (much like the Mats discussed above), and via diffusion to other strata impact the metabolism of the entire column. For example, the top layer supports phototrophs that produce carbon and oxygen, which drives a second layer that aerobically respires carbon and oxygen, resulting in a third layer that is anaerobic and typically uses alternate electron acceptors such as nitrate or sulfate. Recent work has begun to show that these complex communities are amenable to quantitative interrogation in the laboratory \cite{pagaling_community_2014,pagaling_assembly_2017}. Of particular note is a study that used centimeter long glass capillaries to assemble stratified communities cultured in the presence of dyes to report pH and redox~\cite{quinn_winogradsky-based_2015}. A simple imaging setup then permitted the acquisition of quantitative spatiotemporal data on community assembly in many replicates. These systems were used to simulate communities in the lung of a cystic fibrosis patient, but an opportunity to extend this work remains and the platform is well-suited to the ensemble approach outlined here. 

Given that metabolic niches are spatially separated in Winogradsky culture systems they are ideal systems for understanding how this stratification self-organizes, their energetics and how this organization depends on the boundary and initial conditions in the system. For example, if the capillary system of Quinn \emph{et al.}~\cite{quinn_winogradsky-based_2015} could be combined with defined nutrient conditions, controlled illumination and quantitative imaging one could ask how the layers (and therefore the metabolite fluxes) depend on community composition. Given the timescale of these experiments (months) working with many replicates in parallel will be key.



\subsubsection*{Materially closed ecosystems}
Closely related to Winogradsky columns are materially closed microbial ecosystems (CES, Fig.~\ref{Fig4}B). Closed ecosystems are hermetically sealed, typically aquatic, microbial communities that have been shown to sustain life with only light as an input for decades in some cases. These `ecospheres' are available commercially (\url{https://eco-sphere.com/}) and have been studied in an academic setting since the 1960s. CES contain photosynthetic microbes, typically algae or bacteria, and either simple or complex consortia of heterotrophic bacteria and predators. When provided with only light these communities self-organize to sustain nutrient cycles and therefore the community itself. CES act as model biospheres since they require nutrient cycling to persist~\cite{rillig_microbial_2019}. In the context of the structure-function problem the salient metabolic property of CES is nutrient cycling, and one question is how the composition of the community determines nutrient cycling rates and persistence. 

Work from a number of groups~\cite{obenhuber_carbon_1988,obenhuber_eucaryoteprocaryote_1984,kearns_measurement_1981,kawabata_synthesis_1995,taub_closed_1974,taub_community_2009} showed that CES containing primarily microbes were tractable model systems and methods for quantifying nutrient cycling were developed~\cite{obenhuber_carbon_1988,taub_community_2009}. More recently, Leibler and Hekstra~\cite{hekstra_contingency_2012} and later Leibler, Frentz and Kuehn~\cite{frentz_strongly_2015} studied population dynamics in a synthetic closed ecosystem comprised of three species using sophisticated microscopy-based methods. These studies revealed remarkably deterministic population dynamics in CES. However, few studies have quantitatively characterized the nutrient cycling capabilities of CES with the notable exceptions of Obenhuber~\cite{obenhuber_carbon_1988} and later Taub~\cite{taub_community_2009}. The early work of Obenhuber showed that complex bacterial communities mixed with photoautotrophic algae or bacteria could sustain a carbon cycle for many months~\cite{obenhuber_carbon_1988}. Inspired by these studies we recently developed a higher throughput method for quantifying carbon cycling that uses low-cost microelectromechanical sensors (MEMS) made for mobile devices to measure small changes in pressure inside a CES~\cite{astacio_closed_2020}. As appreciated by Obenhuber and Folsome, changes in pressure reflect carbon cycling because oxygen is lower solubility in water than CO$_2$. When photosynthetic microbes fix CO$_2$ they produce oxygen and the pressure rises. When bacteria respire carbon the opposite happens and one can quantify carbon cycling by measuring pressure oscillations during light-dark cycles. We constructed CES using bacterial communities derived from soil samples combined with a domesticated strain of the alga \textit{Chlamydomonas reinhardtii}. We found that taxonomically diverse CES stably cycled carbon for as long as six months. Metabolic profiling of these communities showed that despite taxonomic variability across replicate CES each community exhibited similar metabolic capabilities in terms of the carbon compounds they could utilize. It remains to be understood how such taxonomically distinct communities can retain this diversity while exhibiting similar metabolic capabilities. To address this question using the approach outlined here would require studying many synthetic communities with varying composition while measuring carbon cycling. 

We propose that CES constitute powerful model communities for understanding nutrient cycling. In the context of the approach outlined here, CES could be used to understand how initial nutrient supply (C, N, P, S etc) controls community structure and cycling. Further, since light is the only source of energy CES can be used to explore how energy availability impacts the structure-function mapping in terms of cycling.

\subsubsection*{Enrichment in defined media}

Recently Sanchez and collaborators performed serial enrichment cultures on complex natural communities in simple defined media containing a single carbon source~\cite{Goldford:2018jf,Estrela:hb} (Fig.~\ref{Fig4}C). In this case, the function of the community is the conversion of glucose to CO$_2$ and biomass. Over a few tens of generations these communities assembled into relatively simple, and predictable consortia comprised of a few cross feeding strains. This convergent structure was typified by characteristic ratio of strains from the Enterobacteriaceae and Pseudomonadaceae families. To explain this conserved structure, strains were isolated from the endpoint of the enrichment experiment. They were then assayed for growth on glucose, and Enterobacteriaceae strains were found to grow faster on glucose than the Pseudomonadaceae. Metabolomic measurements (mass spectrometry) revealed that Enterobacteriaceae strains excreted intermediates such as acetate, which Pseudomonadaceae were observed to consume rapidly. Therefore, similar to the polysaccharide particle experiments discussed above~\cite{datta_microbial_2016}, these experiments revealed a primary carbon degrader that rapidly consumed the supplied carbon source but released secondary metabolites (in this case acetate) that supported the growth of other strains. Very recent work suggests that this type of cross-feeding may arise from stress induced release of nutrients that arises due to serial dilution~\cite{amarnath_stress-induced_2021}. In these communities the salient metabolic property of the system is carbon degradation. These studies have already revealed much about how the structure of these communities impacts the carbon degradation (e.g., the role of secondary consumers). It would be interesting to more fully elucidate the mechanistic basis of this cross-feeding in the service of understanding how the genotypes of each strain present determine their uptake and release of carbon compounds. Hopefully these insights would allow us to understand how the cross-feeding depends on available carbon compounds and environmental parameters such as pH. These generalizations could prove insightful in natural contexts such as soils where carbon degradation is a critical phenomenon for climate change.  

\subsubsection*{Bridging structure and function with synthetic communities}
In recent years a number of studies have attempted to bridge structure and function of wild communities by performing experiments on synthetically assembled communities of natural isolates. This approach offers an opportunity to capture much of the substantial taxonomic and genomic diversity of natural communities within a setting where environment and composition can be controlled, and function can be measured accurately.

A handful of studies have explored how interactions between strains in a community affect carbon utilization. In part, these studies attempt to determine how prevalent different types of interaction between strains are (e.g., mutualism, parasitism, etc.), and how these interactions vary based on community composition and the identity of the carbon substrate provided. Using strains isolated from tree-associated environments, Foster and Bell~\cite{Foster:2012kl} assembled synthetic communities and measured CO$_2$ evolution during growth on a complex medium as a metric for community productivity. The measured values of productivity were compared to a “non-interacting” null prediction obtained by summing the productivities of the constituent strains grown in monoculture. What was observed in the vast majority of pair cultures and higher-order communities was consistent with resource competition rather than mutualism, suggesting that competitive interactions for carbon utilization dominate in natural communities. This implies that the carbon utilization of a community should saturate with the diversity of a community, which is precisely what is shown in a more recent study by Yu et al.~\cite{Yu:2019dw}. In this study a diverse ensemble of communities with varying strain composition and diversity was generated from a seawater inoculum via serial dilution and dilute-to-extinction approaches. Cell density, protein concentration, and CO$_2$ evolution were measured along with community diversity (via 16S metagenomic sequencing). These community function measurements show saturating behavior as a function of taxonomic richness. Abundant strains were isolated to disentangle the effects driven by individuals and effects driven by interactions. Interactions increase and saturate with diversity, suggesting both competition and complementation increase simultaneously with diversity. 

Additional recent studies have added nuance to the picture of what interactions are prevalent in carbon-degrading communities~\cite{Kehe:2019in, Kehe:2020ju}, finding a high prevalence of parasitic and mutualistic interactions within communities of isolates that are consistent with cross-feeding. These studies leverage a high-throughput droplet microfluidics platform to perform combinatorial community assembly and culture in multiple different carbon source conditions, and the growth of community constituents was measured using fluorescent labelling. In the first study~\cite{Kehe:2019in}, a high prevalence of competitive interactions was observed, particularly for communities and carbon sources where the constituent strains all showed strong growth on the carbon source in monoculture. However, positive interactions were frequently observed in cases where one strain grew poorly on a carbon source in monoculture, consistent with that strain growing in a community due to cross-feeding of metabolic intermediates. The second study~\cite{Kehe:2020ju} broadened this observation by focusing on communities comprising strains from two taxonomic orders, Enterobacterales and Pseudomonadales. Again it was observed that positive interactions were common in communities where one of the constituent strains grew poorly by itself on a given carbon source. It is likely that the positive interactions were generated by cross-feeding of overflow metabolism intermediates~\cite{amarnath_stress-induced_2021}. These positive interactions likely arise from the same mechanism discussed above in the enrichment culture experiments of Sanchez and colleagues~\cite{Goldford:2018jf,Estrela:hb}. Altogether, these results indicate that community carbon degradation can be a relatively simple function of the taxonomic structure of a community. Linking these relationships to genomic structure remains to be accomplished.


Another study by our group~\cite{Gowda:2021ve} set out to explicitly identify the genomic attributes of a community that are predictive of community function. Using a statistical ensemble of isolates that perform denitrification, a process of anaerobic respiration involving the reduction of oxidized nitrogen compounds, we first mapped the genotypes of individual isolates to the precisely quantified kinetics of nitrate and nitrite reduction, which were parameterized using a consumer-resource modeling approach. We used a regularized linear regression approach to predict nitrate and nitrite reduction kinetics from the presence and absence of denitrification-pathway genes possessed by each isolate. We then assembled communities of these isolates and determined that resource-competitive interactions are prevalent and predictable from single-strain kinetics via the consumer-resource model.  Thus we inferred that the conserved properties of metabolic genes allow the prediction of community-level function. This study shows that synthetic communities comprised of natural isolates, combined with statistical approaches, can yield insights into the mapping from gene content to metabolite dynamics. It remains to be seen if this approach can be applied to more complex communities. 

Our work in Gowda \emph{et al.}~\cite{Gowda:2021ve} points to the idea that focusing on the genomics and ecology of specific metabolic processes can be a powerful approach. In particular, this study leads us to believe that denitrification offers a remarkable system for the quantitative interrogation structure and function both using the statistical approach outlined here, but also for detailed physiological studies of specific metabolic processes. The advantages of denitrification include the fact that the taxa that perform the process are easily isolated and grow well in the laboratory~\cite{lycus_phenotypic_2017}, the metabolites can be quantified in high-throughput~\cite{Gowda:2021ve}, the molecular genetics of denitrification are well-understood~\cite{zumft_cell_1997} and the process has been characterized in the wild to some extent~\cite{tiedje_denitrification_1983}. These opportunities have spawned several compelling recent studies~\cite{lilja_segregating_2016,lilja_substrate_2019,goldschmidt_metabolite_2018}, including a study of the role of carbon source identity in driving denitrification in wild communities~\cite{carlson_selective_2020}. 

\section*{Experimental advances and opportunities to enable the ensemble approach}
\label{S:4}

Having reviewed a variety of model communities for undertaking an ensemle approach to the structure function problem we now turn our attention to experimental methods to study these systems. We focus here on culturing and isolation methods as well as analytical techniques for measuring metabolites. We do not review sequencing approaches that have been discussed in detail elsewhere~\cite{hugerth_analysing_2017}.

\subsection*{Higher throughput/targeted isolation techniques}
A vast majority of microbes that are known to exist in nature remain uncultured in the laboratory. The staggering difference between the number of cells counted from microscopy and those obtained on agar plates was discovered in the early 19th century \cite{amannDirekteZahlungWasserbakterien1911} and was dubbed as the ‘the great plate count anomaly’ in 1985 by Staley and 
Konopka \cite{staleyMeasurementSituActivities1985}. Developments in sequencing techniques have further widened the gap between the number of cultured and uncultured bacteria. The causes for microbial uncultivability include requirement of growth factors present in the natural environment, slow growth, need for interspecies interactions and transitions to dormancy. Development of isolation techniques that overcome these drawbacks is important to capture the high microbial diversity that exists in the wild. Such techniques will aid in the construction of synthetic communities with high genotypic and phenotypic diversity and hence benefit the study of structure-function problem.

Recently developed isolation techniques that offer some advantages over the conventional approach of plating on agar include culturomics, microdroplets and diffusion chambers. In culturomics, communities are tested for growth in a multitude of media conditions using high-throughput techniques, followed by subjecting the communities to mass spectrometry and sequencing~\cite{lagierMicrobialCulturomicsParadigm2012,seng_ongoing_2009}. 
By performing MALDI-TOF (see below) mass spectrometry directly on colonies, bacterial taxa can be identified with high fidelity~\cite{seng_ongoing_2009}. In case MALDI-TOF fails to identify taxa, 16S sequencing is used. The use of mass spectroscopy combined with sequencing facilitates accurate, rapid, and comprehensive strain identification. In a recent study~\cite{lagierCulturePreviouslyUncultured2016}, the culturomics approach was shown to be very successful in increasing the number of species isolated from the human gut (by \char`\~two fold).

A higher-throughput method involves encapsulating cells from natural communities in gel microdroplets (GMDs) made of agar. The GMDs are then incubated in growth chambers flushed with low nutrient media. Following this, GMDs with microcolonies are sorted using flow cytometry and individual GMDs are subsequently transferred into microtiter plate wells containing rich organic medium for biomass enrichment and isolation.\cite{zenglerCultivatingUncultured2002,zenglerHighThroughputCultivation2005} In this technique, the porous nature of the GMDs facilitates exchange of metabolites between droplets during the incubation. As a result, strains that require metabolites produced from resource-mediated interspecies interactions can be isolated using this technique.

Diffusion chambers work by culturing communities in chambers exposed to their native environments through porous membranes~\cite{kaeberleinIsolatingUncultivableMicroorganisms2002, bollmannIncubationEnvironmentalSamples2007,chaudharyDevelopmentNovelCultivation2019}. Thus, the setup allows for the growth of microbes that require growth factors present in their natural environments and/or produced from native community interactions. Significant developments improving the throughput of this isolation method include the isolation chip~\cite{berdySituCultivationPreviously2017a} and the Hollow-Fiber Membrane Chamber (HMFC)~\cite{aoiHollowFiberMembraneChamber2009}.

In addition to these non-targeted isolation techniques, targeted isolation techniques have been recently attempted. These involve designing the isolation methods to target desired phenotypes (e.g., antibiotic resistance or sporulation~\cite{browneCulturingUnculturableHuman2016}). One recent study successfully isolated desired cell types using fluorescently labeled antibodies against predicted cell surface proteins combined with flow cytometry for cell sorting~\cite{crossTargetedIsolationCultivation2019}.

Overall, both the available targeted and stochastic isolation techniques have proven to be useful for isolating previously unculturable bacteria. Hence, these techniques may prove valuable for generating ensembles of bottom-up assembled microbial communities.   

\subsection*{High throughput experimental platforms}
Our proposed ensemble approach for studying structure-function relationship in microbial communities requires creation of many replicate communities. Hence, high throughput culturing platforms are critical for its implementation. 

A majority of high throughput experimental platforms so far have been droplet-based or microfluidic-based devices. One such recently developed device is 'kChip', a microfluidic platform that facilitates combinatorial construction of microbial communities~\cite{blaineyMassivelyParallelOnchip2018}. A study involving synthetically constructed microbial communities on kChips successfully identified sets of strains among 19 soil isolates that promotes growth of model plant symbiont \emph{Herbaspirillum frisingense}, by screening $\sim$\num{100000} multispecies communities~\cite{keheMassivelyParallelScreening2019}. Though kChip is a high throughput platform, it only enables bottom-up construction of microbial communities that requires isolation of microbes prior to the experiments. Additionally, only metabolic functions with optical readouts can be assayed, as physical access to the microdroplets at this scale is not feasible. 

Another similar microfluidic platform that enables parallel co-cultivation of microbial communities was developed by Park \emph{et al.}~\cite{parkMicrodropletEnabledHighlyParallel2011}. Their platform was able to successfully  detect pairwise symbiotic interactions in communities when the symbionts were in as low an abundance as 3 percent of the total population. Here again, only optically detectable metabolic properties can be measured, but the device enables top-down construction of microbial consortia through random compartmentalisation of community members. Though this was not a structure-function study per se, inclusion of automated droplet sorting and characterization of communities in the retrieved droplets can easily enable structure-function studies. In fact, this was achieved in a more recent study by Terekhov \emph{et al.}, where microbes conferring antibiotic resistance in the oral microbiota of Siberian bears were identified~\cite{terekhovUltrahighthroughputFunctionalProfiling2018}. This was done by functional profiling of the encapsulated communities from the oral microbiome that suppressed the growth of the pathogen \emph{Staphylococcus aureus}. 

  
From the aforementioned studies, it can be said that the choice of the experimental platform largely depends on the nature of the study. Some existing platforms support a top-down approach whereas others support a bottom-up approach. Further, the methods of determining structure and function can differ across platforms. There is room to improve these methods to incorporate other analytical techniques for measuring metabolites. For example, if large scale culturing platforms could be combined with spectroscopic or automated mass spectrometry methods, this would enable the rapid construction of large quantitative datasets.

\subsection*{Measuring metabolite dynamics}

Metabolites, unlike nucleic acids, require distinct analytical techniques depending on the metabolite of interest. Here we review the available methods, their applicability and opportunities for improving these methods for microbial consortia. See Table~\ref{MethodsTable} for the specific strengths and limitations of each technique discussed here.

\begin{table}[h!]
\centering
 \begin{tabular}{||c c c c c||} 
 \hline
 Method & Sensitivity & Specificity & \shortstack{Range of \\ applicability} & Throughput \\ [0.5ex] 
 \hline\hline
 NMR & low & high & high & low \\ 
 Mass Spectrometry  & high & high & high & low/moderate \\
 Infrared/Raman  & moderate & high & high & high \\
UV/Vis  & moderate & low & low & high \\
Targeted assays  & moderate & high & low & high \\
 \hline
 \end{tabular}
 \caption{Comparison of methods for measuring metabolites in microbial communities.  Sensitivity refers to the minimum detectable concentration. Specificity refers to the ability of the assay to detect a specific metabolite. Range of applicability refers to the diversity of metabolites that can be detected with the technique. Throughout is the number of measurements that can be made in parallel.\label{MethodsTable}}
\end{table}

\subsection*{Nuclear magnetic resonance spectroscopy}

A number of high quality textbooks describe the fundamental physics~\cite{slichter_principles_1990} and chemistry~\cite{levitt_spin_2008} of nuclear magnetic resonsance (NMR) spectroscopy. Here we give an intuitive explanation of the basis of this technique and go on to the practical applications to measuring metabolites in microbial communities.

NMR spectroscopy exploits the spin magnetic moment of atomic nuclei such as hydrogen, carbon, and nitrogen to characterize chemical structure. In an applied magnetic field, nuclei behave as weak magnets, collectively aligning with the field. The collective alignment of the nuclear magnetic moments can then be manipulated with applied electromagnetic fields in the radio frequency (MHz) and detected as emitted fields in the same spectral region. The small magnetic moments of nuclei cause them to emit very weak radiation meaning that relatively high concentrations of metabolites of interest are necessary for detection.  Nuclei experience minuscule changes in the local magnetic field due to their local chemical context, resulting in what are termed ``chemical shifts.'' For example, a proton in hydrogen on an alkane (e.g., methane) will emit a distinct radio frequency (its resonance frequency) from one in an aromatic hydrocarbon (e.g., benzene). These small changes in emitted radio frequency fields are of the order of parts per million (ppm). Typical resolution of modern instruments are a fraction of a ppm and depend on the field strength of the spectrometer and technical details of the detection scheme. State-of-the-art spectrometers (operating at \num{600}MHz and above) are widely available at core facilities.

For metabolomics the two most common types of NMR are proton ($^1$H) and carbon NMR, each of which has both advantages and disadvantages. First, the advantages of proton NMR are (1) rapid acquisition due to the relatively high signal-to-noise ratio in proton spectra, (2) the fact that no isotopic labeling is necessary, and (3) the ability of the technique to detect a broad range of relevant organic compounds. One disadvantage is the fact that spectra from mixtures of unknown metabolites are complex, often containing hundreds of peaks corresponding to the many different compounds present. As a result, it can be challenging to detect the presence or absence of specific metabolites via proton NMR. Further, water contributes a broad and strong solvent signal in the middle of the relevant range of chemical shifts. There are two main routes to removing this signal: (1) drying the sample and replacing the water with D$_2$O and (2) using clever pulse sequences to decouple the water signal from the metabolite signals~\cite{mckay_how_2011}. The former requires specialized equipment and increases the cost and reduces throughput. Therefore, it is recommended to use decoupling. The fundamental physics of how this decoupling works is beyond the scope of this review, but it is recommended to use 1-D NOESY (Nuclear Overhauser Effect Spectroscopy) to isolate metabolite signals from water~\cite{emwas_nmr_2019}. The approach is robust, widely applied, and requires no sample processing to be done. It should be noted that because of variation in technical specifications between instruments, performing measurements on a single spectrometer across samples is key to maintaining reproducibility of measurements~\cite{mckay_how_2011}.

Carbon NMR in contrast, detects signals from the magnetic moments of carbon nuclei. As with protons, the chemical context of the carbon nucleus gives rise to chemical shifts in the resonance and this permits the disambiguation of carbon nuclei in different compounds. One advantage of Carbon NMR is that it does not require suppression of water signals. The main drawback to carbon NMR is sensitivity. The dominant isotope of carbon ($^{12}$C, \SI{99}{\percent} prevalence) is not NMR active, while $^{13}$C is NMR active, but present at about \SI{1}{\percent} natural abundance. This means that most nuclei do not contribute to the observed signal. Second, $^{13}$C has a magnetic moment that is roughly \num{4}-fold lower than $^1$H reducing the signal-to-noise ratio. These two considerations imply that acquiring carbon spectra requires extensive averaging and can take hours for a single sample. However, the low isotopic abundance of $^{13}$C can be overcome by using $^{13}$C labeled compounds as nutrients, with order of magnitude increases in signal-to-noise. Unfortunately, these compounds are expensive (hundreds of dollars per gram) increasing costs. 

Carbon and proton NMR are the two most commonly applied metabolomic profiling techniques for microbial communities. Typical studies range from targeted detection of a single metabolite in a well-defined community~\cite{andrade-dominguez_eco-evolutionary_2014} to untargeted profiling in very complex consortia such as anaerobic digesters\cite{gonzalez-gil_nmr_2015}. Here we present a few examples of NMR based measurements of metabolite dynamics in communities as case-studies that might be more broadly applicable.

One compelling approach taken by Date \emph{et al.}~\cite{date_new_2010} and Nakanishi \emph{et al.}~\cite{nakanishi_dynamic_2011} is to combine NMR based metabolite measurements in time with quantification of abundance dynamics. Date \emph{et al.} use $^{13}$C labeled glucose to initiate growth in fecal microbiota~\cite{date_new_2010} (note that $^{13}$C is a stable isotope). The authors then performed time series of abundance dynamics and carbon NMR measurements. Given labeled glucose as the sole carbon source, the authors could track the dynamic production and consumption of $^{13}$C labeled compounds as the glucose was converted to organic acids in time. The authors could then (crudely, given the electrophoresis methods at the time of the study) classify the community into primary and secondary degraders. The compelling aspect of this study is the potential to statistically correlate large scale variation in the community structure with metabolite dynamics. One could imagine a similar experiment with amplicon sequencing-based abundance dynamics measurements. This approach would be especially powerful for looking at the community level response to carbon fixation by autotrophs in systems like mats or CES. In these situations initiating a community with $^{13}$C labeled bicarbonate as the sole carbon source would allow the direct measurement of carbon flux from autotrophs to associated heterotroph communities. Since NMR relies on magnetic fields and emitted RF signals, it is non-invasive and could be applied to ongoing experiments (e.g. CES) without invasive sampling. 

We conclude with a brief note on throughput. NMR spectrometers are large, expensive machines that rely on superconducting magnets to apply large magnetic fields. Running parallel experiments on NMR machines is therefore prohibitive. Throughput is achieved by using robotics or fluidic systems to automatically load samples into the spectrometer. Such experiments typically achieve a throughput of order \num{100} samples per day~\cite{macnaughtan_high-throughput_2003,soininen_high-throughput_2009}. Increasing NMR throughput by a factor of 10-100 would constitute a major advance.

Finally, despite the current limitations, there is a revolution underway in quantum sensing systems from superconducting quantum interference devices (SQUIDs)~\cite{mcdermott_liquid-state_2002} to spin-based magentic field sensors,  impurities in diamonds (nitrogen-vacancy (NV) centers)~\cite{cujia_tracking_2019} or force measurements\cite{kuehn_advances_2008}. SQUIDs enable ultra-low field NMR, obviating the need for large expensive magnets, and NV-centers enable high sensitivity magnetic resonance detection at the single-molecule level. The applications of these technologies to metabolic function in microbial communities await future discovery, but one can imagine massively parallel NMR measurements or \emph{in situ} detection of metabolites in complex settings.

\subsection*{Mass spectrometry}

Mass spectrometry is the most widely used platform for metabolomics and several good reviews of the methodology are available~\cite{beale_review_2018,mastrangelo_sample_2015,dettmer_mass_2007,raftery_mass_2014,alseekh_mass_2021,Jemal2000}. Therefore, our discussion of mass spectrometry will be limited, but we include it as an important point of comparison with the other techniques discussed in this section.

Mass spectrometry ionizes the molecules in a sample, using a variety of different methods, and then accelerates the charged molecules using an electric field. The beam of ions is then passed through a magnetic field that (via the Lorentz force law) results in a force on each ion that depends on its mass to charge ratio. The result is a physical separation of ions in space in proportion to the mass-to-charge ratio. A huge number of variations exist on this basic theme, including measurements of time of flight (TOF) and quadrupole mass filters that apply oscillating fields to the ion beam. The reviews cited above contain detailed discussions of the type of ionization and detection methods that are best suited for metabolomic applications.

In the context of metabolomics, where samples are often of considerable chemical complexity, mass spectrometry is almost always preceded by either gas or liquid chromatography to separate compounds and therefore increase the specificity and sensitivity of downstream mass spectrometry. As a result, gas chromatography-mass spectrometry (GC-MS) and liquid chromatography-mass spectrometry (LC-MS) are the two most widely used forms of mass spectrometry for metabolomics. GC-MS is restricted to volatile compounds typically of molecular weight less than 600 Da~\cite{beale_review_2018}, while LC-MS applies to a broader spectrum of metabolites.

Mass spectrometry has significantly higher sensitivity that NMR, and can achieve single molecule sensitivity~\cite{robertson_single-molecule_2007}, although this is not routine. This advantage is especially crucial for measuring metabolite fluxes in cells and communities where concentrations are often low (micromolar and below)~\cite{bennett_absolute_2009,basan_universal_2020}. This combination of sensitivity and the ability to distinguish broad classes of compounds has contributed to the widespread usage of MS-based metabolomics methods~\cite{aiyar_antagonistic_2017}. 

The throughput of these techniques is significantly lower than the optical methods discussed below and comparable, at present, to NMR. So it is routine to run $\sim$\num{100} samples over the course of a day or two (see~\cite{Jemal2000} or a review). Some robotic systems have been developed to automate the sampling and analysis process~\cite{molstad_robotized_2007}. However, making mass spectrometry measurements on thousands or tens of thousands of samples, while feasible, is costly and slow. Given the remarkable sensitivity, specificity and broad applicability of the technique especially for untargeted measurements of metabolite pools, it would represent a major advance if mass spectrometry could be routinely applied to thousands of samples in parallel. Here, we mention a handful of notable studies that leverage mass spectrometry to quantify metabolite dynamics in microbial communities. The goal is not to present an exhaustive list but simply to point the reader towards some representative studies. 

Amarnath \emph{et al.}~\cite{amarnath_stress-induced_2021} used untargeted metabolomics to study the metabolites that are exchanged between two strains of bacteria in a serial dilution experiment. The authors revealed a broad spectrum of metabolites excreted by one strain in response to stress. These excreted compounds facilitated cross-feeding between the two strains. Shi \emph{et al.}~\cite{ShiLCMS} use LC-MS to study metabolite exchange in a fungal-bacterial community. A statistical analysis of the LC-MS data shows that the metabolites excreted are distinct for co-cultures and mono-cultures.

Mass spectrometry is widely used to study the degradation of compounds from pharmaceuticals to soil contaminants~\cite{PIEPER20104559,THELUSMOND20161241}. For example, a common environmental contaminant are polycyclic aromatic hydrocarbons (PAHs), which are routinely degraded by bacterial consortia. The breakdown of these compounds in time is typically interrogated by GC- or LC-MS~\cite{luan_study_2006}. %

Mass spectrometry is widely applied to food-related microbial communities. In these cases the untargeted nature of mass-spectrometry is important as the compounds of interest (e.g., for flavor) are typically unknown. For example, GC-MS has been used to identify starting components in traditional Cambodian rice wine \cite{ricewineGCMS}. Similarly, it was used in fermentation of red peppers to investigate changes in bacterial and fungal communities and volatile flavor compounds \cite{XUredpepper}, and high throughput GCMS has been used to correlate metabolites with taxonomic structure in kimchi fermentation \cite{PARK2019558}, glutinous rice wine \cite{HUANG2019593}, the liquor \textit{Daqu} \cite{JIN2019422} and pickled radishes \cite{RAO2020108804}.



\subsection*{Infrared spectroscopy}
Infrared spectroscopy detects absorption and emission of photons in the infrared region of the electromagnetic spectrum and characterizes molecular vibrations. As with NMR, the resonance frequency of a molecular vibration depends strongly on molecular structure. This dependence affords IR spectroscopy its chemical specificity. Infrared spectroscopy offers perhaps the best combination of sensitivity, specificity, high-dimensional characterization of complex metabolite pools and throughput. There are two commonly used methods for measuring infrared spectra that differ in their fundamental physical mechanisms: (1) IR absorption and (2) Raman spectroscopy. 

Infrared absorption involves passing light in the infrared range ($\sim$\SI{2.5}{\micro\meter} to \SI{10}{\micro\meter} wavelengths or \SI{1000}{\per\centi\meter} to \SI{4000}{\per\centi\meter} wavenumbers) through an aqueous or gas phase sample and measuring the absorption. Compounds containing different chemical bonds absorb light at different frequencies and the resulting spectrum can provide extensive information on the chemical composition of the sample. As with proton NMR, a major downside of absorption spectroscopy is the broad absorption of water in the informative region of the spectrum (around \SI{3200}{\per\centi\meter} and \SI{1600}{\per\centi\meter}), which can limit the information for aqueous samples without cumbersome drying. Simple dispersive spectrometers that shine a narrow band of wavelengths through a sample have limits on sensitivity, spectral resolution and the duration of acquisition. These limitations can be overcome using Fourier Transform infrared spectroscopy (FTIR), which uses broadband excitation and an interferometer to rapidly acquire spectra in specific spectral bands, and this is the most commonly applied technique. Plate readers that perform FTIR measurements en masse on microtiter plates are available and can acquire data from both liquid and solid phase samples.

In contrast to FTIR or dispersive IR spectroscopy, Raman spectroscopy measures molecular vibrations using photons in the visible portion of the spectrum. When visible photons interact with a sample, most are scattered with the same energy as the incident photon (Reyleigh scattering). However, with low probability, incident light undergoes inelastic scattering and in the process photons are emitted from the sample with either slightly lower (Stokes) or higher (anti-Stokes) energy than the incident radiation. These small changes in the emitted photon wavelength correspond to the molecular vibrations in the sample. For reasons beyond the scope of this review, Raman spectroscopy does not suffer from broad band absorption from water, making it especially attractive for microbial communities in the aqueous phase. However, due to the inefficiency of inelastic photon scattering, Raman spectroscopy requires high laser power and the resulting heating can be a problem for biological samples, a limitation that can be overcome by techniques such as resonance Raman spectroscopy or surface enhanced Raman scattering. However, these methods are not yet routine for metabolomic profiling. Raman spectroscopy can be performed on bulk samples using plate readers, or integrated with a microscope for localized measurements. More recently, Raman spectra can be acquired via flow cytometry~\cite{suzuki_label-free_2019,nitta_raman_2020}. These platforms enable much higher throughput than is now standard by NMR or mass spectrometry. 

FTIR and Raman spectroscopy have proven to be powerful methods for interrogating cellular physiology at the single-cell level when combined with microscopy (see ~\cite{hatzenpichler_next-generation_2020} for a recent review). Remarkably, Raman spectroscopy signals can be used as fingerprints to demarcate cells of one species in different growth states~\cite{escoriza_studying_2006}, or different taxa at the strain, species and genus levels~\cite{rosch_chemotaxonomic_2005,harz_micro-raman_2005}. A recent study showed that the global transcriptional profile of yeast and bacteria could be predicted via a linear model from single cell Raman spectra\cite{kobayashi-kirschvink_linear_2018}. This success owes to the high-dimensional nature of Raman spectra, which are often challenging to interpret in terms of individual peaks but are rich in information that can be decoded statistically.

Despite the power of infrared spectroscopies for chemical characterization, they have seen comparatively little use in the context of communities of microbes. We regard this as a missed opportunity, and suggest that these methods could and should be used more broadly. One of the limitations of the technique is the challenge of assigning specific peaks to specific compounds. As Kobayashi and co-workers have shown, this limitation can be overcome by using simple statistical methods to map infrared spectra to other cellular properties \cite{kobayashi-kirschvink_linear_2018}. The approach is to measure Raman or IR spectra on a set of samples and then use a lower throughput technique such as LC-MS to measure the absolute concentration of a metabolite of interest. A combination of dimensionality reduction and regression can then be used to map LC-MS data to infrared spectra. This approach has been used to track substrate concentrations in time in monocultures~\cite{paul_towards_2016} and phenol degredation in complex communities~\cite{wharfe_fourier_2010}.

The advantages of Raman spectroscopy, and to a lesser extent FTIR, over mass spectrometry and NMR are the potential for the rapid acquisition of high-dimensional characterization of metabolite pools. The complexity of the resulting spectra is similar to proton NMR, and therefore is perhaps most useful for statistically characterizing differences between community metabolite profiles. Such complex spectra can then be used either to measure specific metabolites via a calibration approach discussed in the previous paragraph, or to demarcate global metabolic states of consortia without concern for specific metabolite levels. 


\subsection*{UV-visible spectroscopy}
We briefly note that simple ultraviolet and visible spectroscopy (UV-Vis) can be used to characterize electronic transitions in compounds of interest for metabolic characterization. These methods can be performed with widely available plate readers, particularly those that are equipped with monochromators rather than filters, which permit excitation and emission to be arbitrarily selected by the user. The main limitation of this technique is the fact that electronic transitions in the visible are restricted primarily to chemical species with delocalized electron density (e.g., conjugated rings such as benzene, tryptophan). As a result, these spectra are low specificity and cannot be used for targeted metabolomics. However, the high throughput of common plate readers facilitates rapid measurements, and the spectra can give coarse characterization of excreted compounds from autotrophs, for example~\cite{tenorio_impact_2017}. Moreover, targeted UV-Vis measurements can be integrated with common fluorescence microscopes and therefore offer the possibility of detecting metabolites in massively parallelized platforms such as droplet microfluidics~\cite{Kehe:2019in,Kehe:2020ju} or in spatially structured communities.

\subsection*{Targeted assays}

In situations where the metabolic function of the community under study is known apriori and restricted to a specific chemical compound a targeted metabolite assay can be powerful. Such assays typically utilize chemical reactions to create an optically active compound in proportion to the concentration of a metabolite of interest. For example, starch can be degraded to glucose enzymatically and then glucose concentration can be assayed via standard methods~\cite{holm_rapid_1986} or iodine can be used to stain starch directly and detect degredation~\cite{fuwa_new_1954}. Similarly, nitrate and nitrite can be detected via the Griess assay~\cite{miranda_rapid_2001}, which utilizes a colorometric reporter (dye) generated via a reaction with nitrite. Such assays can easily be performed in 96-well plates facilitating high throughput \cite{Gowda:2021ve}. These measurements can be powerful for studying specific metabolic processes in communities. However, typically the chemistry involved is not easily automated nor are the conditions of the reaction biocompatible. So, such measurements are made offline after sampling and are challenging to automate \emph{in situ}.



\section*{Quantifying structure}

We briefly outline the main ways in which community structure can be quantified. As mentioned above a suite of next generation sequencing technologies are capably of quantifying community structure on multiple levels. For example, amplicon sequencing of the 16S rRNA gene uses PCR to amplify this universally conserved ribosomal subunit and then uses the number of reads mapping to sequence variants (amplicon sequence variants, ASVs) as a proxy for the relative abundance of each taxon. This widely used method has many well documented downsides including variation in the copy number of the 16S gene across taxa, PCR bias and the challenge of associating taxonomy with metabolic capabilities of each strain \cite{Callahan2016}. Despite these shortcomings, amplicon sequencing does permit rapid and high throughput characterization of the community composition, and methods exist for inferring metabolic capabilites of strains from 16S gene sequence alone~\cite{douglas_picrust2_2020}.

In contrast, shotgun metagenomic sequencing amplifies all genomic DNA in a sample. These data enable the characterization of the gene content of an entire community by annotating reads. Metagenomics gives a much more complete picture of the genomic structure of a community but several technical hurdles limit this approach. First, annotating reads mapping to genes remains a challenge and roughly \num{30}-\num{50}\% of the open reading frames are annotated, leaving much of the genomic content unclassified in terms of function. Second, assigning annotated genes to specific taxa within the community and inferring their relative abundances remains a hard problem. In particular, assembling reads into genomes (metagenome-assembled genomes, MAGs) has been performed but the quality of these assemblies remains hard to assess. Applying cutting edge machine learning methods is likely to improve this process~\cite{nissen_improved_2021}. Another method to analyze shotgun metagenomics data is to use a database of reference genomes as templates to recruit reads from a metagenome. The upside of this approach is the ability to reliably detect variation at the level of single nucleotides~\cite{garud_evolutionary_2019}, and reliably assemble genomes from metagenomes. The cost of course is that the method misses any diversity that is not present in the reference genome database. Despite these challenges, metagenomics perhaps gives clearest picture of the genomic structure of a community as a whole.

Transcriptional profiling of entire communities is also feasible via RNA-sequencing based methods~\cite{antunes_microbial_2016,zhang_metatranscriptomics_2021}. Despite the potential predictive power of knowing which genes in a community are transcriptionally active, these measurements have been applied much less widely than taxonomic amplicon or shotgun metagenomic methods. However, as the costs of sequencing continue to fall, it remains a compelling proposition to use metagenomics and transcriptional profiling on the same samples. We propose that such measurements could very well lead to deeper insights into the community structure and function by potentially simplifying the picture. For example, transcriptional profiling could reveal which collective components of the metagenome are inactive and therefore could potentially be left out of a predictive framework.

\section*{Learning from data: function from structure}
\label{S:5}

Equipped with measurements of metabolites (either dynamically or at a fixed point in time) and some characterization of the community structure, we are then left to ask what to do with these data. In reality, the answer to this question is an empirical one that depends on the structure the data, the model system under study and the precise question being posed. We offer no pipeline or prescription for how best to proceed, but instead offer a few suggestions and examples and point out some important technical pitfalls. Our intention is to suggest some approaches to learning the structure-function mapping from these data and to leverage the results for predicting community metabolism. As with most data analysis tasks, simple is better. Using the latest methods in machine learning or dimension reduction may be tempting but it is almost always better to explore the data with methods that are simple to interpret and straightforward to implement.

\begin{figure}
\centering\includegraphics[scale=0.8]{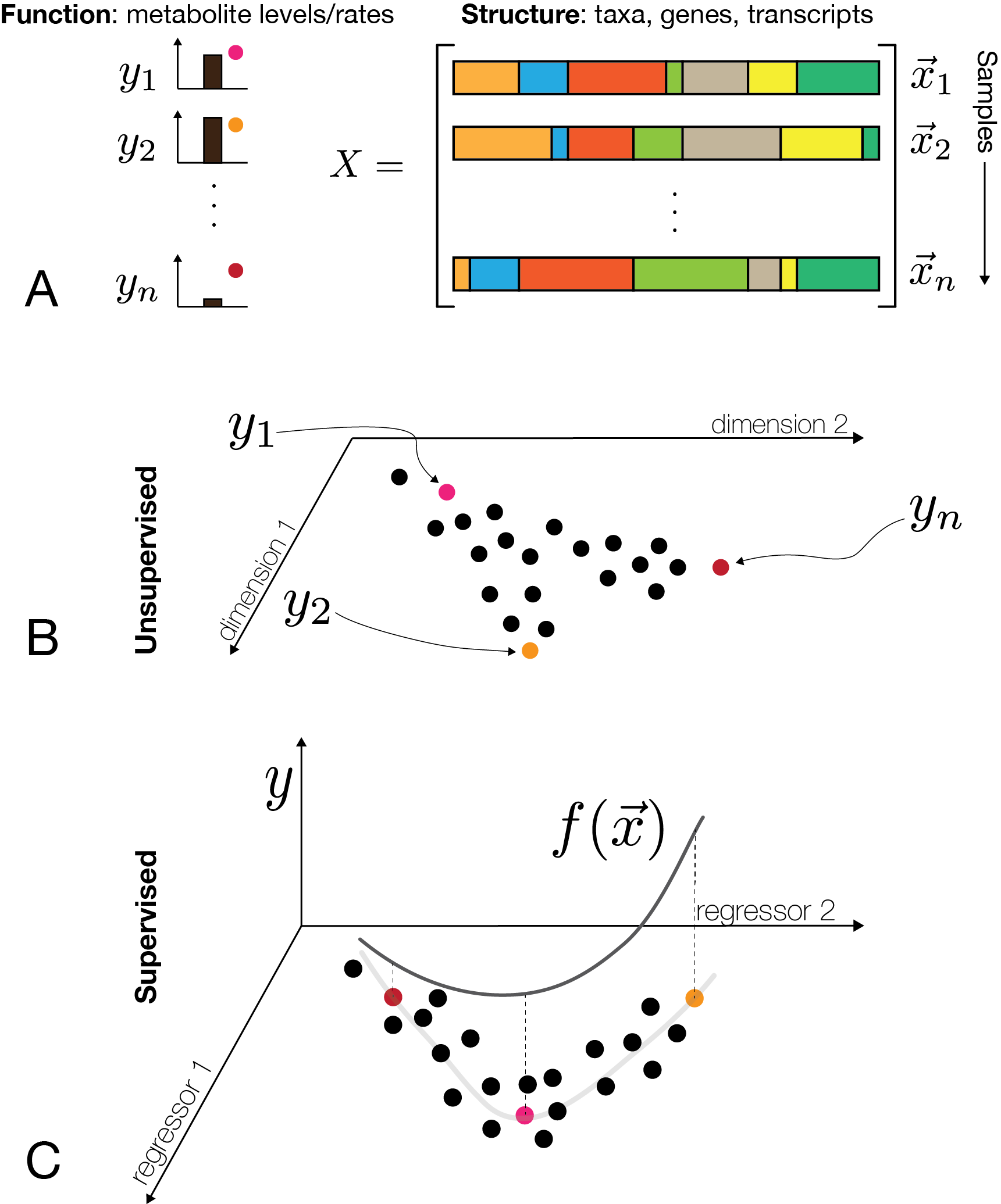}
\caption{\textbf{Learning the structure-function map from data.} (A) Hypothetical structure-function data from an ensemble of $n$ communities. $y$ denotes a metabolite measurement, either level or rate that could also be dynamic. Colored dots correspond to data points in (B,C). $X$ denotes a matrix of $n$ rows each denoting a single community in an ensemble. The columns denote the relative abundances (colored bars) of taxa, genes in the metagenome or transcripts in the metatranscriptome. (B) An unsupervised approach where dimensionality reduction is applied to $X$ yielding a lower dimensional representation of community structure that is then associated with communities of differing function. (C) A supervised approach where the function $f(\vec{x})$ is learned for mapping structural variation to functional variation. Regressors denote independent variables in a lower dimensional representation of $X$ that provides good predictive power of $y$.}
\label{FigData}
\end{figure}

Typically any sequencing based characterization of community structure will be high dimensional. For example, 16S amplicon sequencing will often yield \num{10} to \num{1000} of taxa per sample, similarly metagenomic data can contain \num{10000} or more annotated open reading frames depending on the complexity of the community. In contrast, assembling an ensemble of more than \num{100} or \num{1000} communities is a huge challenge even for the highest throughput methodologies. As a result, we are almost always in the limit of a small number of data points (communities) and large number of variables (taxa, genes, transcripts etc). In this regard, predicting functional properties from these structural data requires reducing the dimensionality of the data describing community structure. For the purposes of the discussion below we define the number of features (genes, taxa, transcripts) as $p$ and the number of communities in a given ensemble $n$ (Fig.~\ref{FigData}A). A sequencing data set can then be described by a matrix $X$ that is $n\times p$ with $n<<p$ in most cases. The rows of this matrix, $\vec{x}_i$ correspond to sequencing data for the $i$th community in the dataset. The entries of this vector are then the number of reads mapping to a specific ASV in a 16S data set or gene in the metagenome of that community. For each of these communities we assume that the data set includes some functional measurement, $y_i$, which may be a dynamic quantity ($y_i(t)$). The goal then is to learn a representation of this functional measurement of the community in terms of the columns of $X$. 


\subsection*{Compositional data and zero counts}

All of the standard sequencing methods for quantifying community structure result in compositional data - that is they do not report the absolute abundances of taxa, genes or transcripts in the sample, but only the relative contribution (e.g., the rows of $X$ are defined only up to an unknown constant). Recently, methods using qPCR or the addition of oligos at known concentrations have been developed to measure absolute abundances via sequencing. However, as yet these methods are not widespread. Therefore, in any analysis we must contend with the compositional nature of sequencing data. Much has been written about this problem~\cite{gloor_microbiome_2017}, and members of the field are now generally aware of the issues that can arise when the compositional nature of the data are ignored.

Briefly, compositional data can, and should, be log-ratio transformed using log ratios of counts. Log-ratio transformations take compositional data from a simplex and map them to real numbers with the properties of a vector space. This transformation therefore permits the application of conventional statistical approaches to compositional data. Typically, this is done via the center-log transform (CLR) or an additive log-transform (ALR)~\cite{Aitchison2000,gloor_microbiome_2017}. Computing log-ratio is not compatible with zeros (e.g., zero abundances of an ASV or gene transcript), a problem that has received a significant amount of attention. A host of methods from adding psuedocounts uniformly to all zeros or using Bayesian approaches to replacing zeros ~\cite{Aitchison1982,love_moderated_2014,mcmurdie_waste_2014}. We urge caution here, as many methods are both ad hoc and can qualitatively impact the results of downstream analyses. We will not review the technical details, but readers should to engage carefully with their data rather than blindly applying existing pipelines for log-transforms and handling zeros. For example, variance decompositions applies to log-transformed data can be dominated by large numbers of taxa or transcripts with zero counts. In this scenario the details of how zeros are handled (e.g., the magnitude of the psuedocounts added to all taxa) can have huge impacts on the variance decomposition.

\subsection*{Dimension reduction: with or without supervision}

The goal is to associate structural components or sets of components with the metabolic function of a community. As discussed above we are almost always in the limit of small numbers of data points. We are therefore forced to consider reducing the dimensionality of the data from $p$ by some form of dimension reduction. Above we suggested that low-dimensionality is a common feature of biological systems from proteins to higher organisms and behavior. Despite the fact that this observation seems to hold quite broadly it is important that we not take low-dimensionality as given in any analysis of a microbial community. In this sense, we must make a principled search for a simpler description of community structure while judiciously considering the possibility that no such description exists or that we are considering the system at the wrong level of organization. For example, in an ensemble of communities with strong functional redundancy, where very distinct taxa perform similar metabolic functions, it may not be possible to find a low-dimensional description of the ensemble in terms of taxa present across replicates in the ensemble.  


There are two ways to go about approaching this problem: supervised and unsupervised dimension reduction (Fig.~\ref{FigData}B-C). We note that for most of the techniques described below the statistical learning textbook from Hastie, Tibshirani and Friedman~\cite{hastie_elements_2016} provides an excellent and readable reference. For a more recent review article we recommend the paper of Mehta \emph{et al.}~\cite{mehta_high-bias_2019} that covers both supervised and unsupervised methods.

\textbf{Unsupervised learning:} In unsupervised dimension reduction we seek a lower dimensional representation of the matrix $X$ without any explicit consideration given to $y$ in selecting low-dimensional features (Fig.~\ref{FigData}B). The idea is to reduce the dimensionality of $X$ and then examine the relationship between this lower dimensional representation of community structure and function, $\vec{y}$. The hope is that some lower dimensional representation of the community captures all of the useful information in the matrix $X$. All of these methods change the basis of $X$ in order to recast community structure in terms of groups of genes or taxa. The dimension reduction arises when a small number of such groups of taxa provide a good description of all communities in the data set. If this is case each community $\vec{x}$ can be represented not in terms of the abundances of $p$ taxa, but instead as a combination of $m<<p$ groups or combinations of taxa. Formally, these methods are based on a matrix decomposition ($X = AB$, with $A$ an $n \times m$ matrix and $B$ an $m \times p$ matrix) subjected to constraints. Dimension reduction comes about if $X$ can be well approximated for $m<<p$. In this case, each community in an ensemble can be represented in terms of just $m$ features. These low-dimensional representations, when they exist, can sometimes dramatically simplify our understanding of complex systems.

The canonical unsupervised methods are variance-based decompositions like principle components analysis (PCA) or the more general version, singular-values decomposition (SVD). These methods find the set of orthogonal directions in the $p$-dimensional feature space that maximize the variance in the data along each direction (eigenvectors of the covariance matrix of $X$). The advantage here is the clean interpretability of these decompositions. Each direction (principle component) is a simple $p$-dimensional vector representing a direction of high data variance. Each data point (community) in the original data set can be represented in the basis of principle components $\vec{x}_i \cdot \vec{p}_1,...,\vec{x}_i \cdot \vec{p}_j,... $ (with $\vec{p}_j$ the principle components). If a small number of principle components capture a substantial fraction of the data variance then $m$ is small and each community can be represented as projections on a handful of $\vec{p}_j$ components.

Many modifications to this basic idea exist. Perhaps the two most notable are independent components analysis (ICA)and non-negative matrix factorization (NMF). ICA finds a lower dimensional representation of $X$ in terms of components that are statistically independent, rather than simply uncorrelated, through an iterative process. The approach is to represent each community $\vec{x}$ as a sum of statistically independent components~\cite{hyvarinen_fast_1997}. In this case $\vec{x}_i = \sum_j a_{i,j} \vec{s}_j$ where the $a_{i,j}$ are the weights that `mix' the independent components $s$ to form the observed data $\vec{x}$. ICA is most often applied to signal separation problems where multiple independent inputs are combined. One can speculate that this perspective may be useful for communities if they possess a modular organization where independent modules within the community are responsible for functionally distinct processes. NMF is another matrix decomposition method that is applicable whenever the entries of $X$ are constrained to be positive (as is the case in sequencing data). NMF decomposes the data as $X \approx WH$ with all entries of $W$ and $H$ positive~\cite{lee_learning_1999}. In this case $\vec{x}_i = \sum_j w_{i,j}\vec{h}_j$ where the $w$ are weights and the $\vec{h}$ are vectors of length $p$. The advantage of this approach is that the columns of $H$, that act as `eigen-communities' contain all positive entries and are therefore interpretable. In contrast, the eigenvectors of a PCA decomposition can contain negative entries, which is not interpretable in the context of $X$ that contains only positive values (abundances or relative abundances). NMF has seen limited application in the microbiome context~\cite{cai_learning_2017}. We have focused here on well-established methods with simple interpretations. We are aware that over the past two decades many new methods have been developed, especially those that can learn low-dimensional representations of highly nonlinear data (e.g., autoencoders~\cite{kramer_nonlinear_1991} or stochastic neighbor embedding methods~\cite{hinton_stochastic_2002}). These methods may be useful in the context discussed here, but we advocate starting with the simpler approaches discussed above before moving on to these methods.  

Regardless of the method of unsupervised learning applied the result is a new representation of community structure ($X$) in a new basis (e.g., principle components, $\vec{p}$, $\vec{s}$, $\vec{h}$). Ideally, $m<<p$ and communities with many hundreds or thousands of genes or taxa can be represented in a much lower dimensional space ((Fig.~\ref{FigData}B). The task then remains to associate this lower dimensional representation with metabolic function ($y_i$). A common approach to this problem is simply to ask which basis vectors correlate with specific metabolic properties of the community. One common approach is to treat the question as a regression problem to predict $y_i$ from the decomposed $\vec{x}_i$ for example by using the projections of each $\vec{x}_i$ along each principle component as independent variables. 

\textbf{Supervised learning:} One major shortcoming of the unsupervised approach is that the low-dimensional representation of $X$ that we learn by unsupervised dimension reduction may not be the best representation of the data in terms of predicting $y_i$. In essence, PCA or NMF may find a low dimensional representation of the data, but there is no reason or guarantee that this representation will be informative of the community metabolic function. Indeed, the unsupervised approach artificially separates the process of finding low-dimensional descriptions of the community and predicting the response variable (metabolic function). A supervised approach overcomes this limitations by performing both dimension reduction and prediction at the same time.

In the supervised approach we seek some prediction of the metabolic function in terms of the structure (Fig.~\ref{FigData}C). Concretely, we would like to estimate $y_i = f(\vec{x}_i)$ for our entire data set. This can be posed as a regression problem where we either posit a functional form for $f(\vec{x}_i)$ or, using more flexible but less interpretable methods like neural networks, learn a mapping from $\vec{x}_i$ to $y_i$ without positing a specific functional form. 

Before discussing how one can approach this problem we would like to clarify the meaning of $f(\vec{x}_i)$. We are proposing learning a statistical map from $\vec{x}_i$ to $y_i$. We are not proposing fitting an explicit ecological model such as a consumer-resource or Lotka-Volterra model to the data. We regard fitting such a model as a much harder proposition than learning a statistical mapping from structure to function. Indeed, a statistical approach to learning $f$ necessarily abstracts away these dynamics that relate $\vec{x}_i$ to $y_i$. We note that two of us recently took a hybrid approach to the problem with synthetic communities that did explicitly model the ecological dynamics~\cite{Gowda:2021ve}. In this case, we used a regression to map gene content to consumer-resource model parameters and then used the consumer-resource model to predict metabolite dynamics in consortia. Remarkably, the approach worked, but the downside is that it requires isolating individual taxa and constructing synthetic communities - a more laborious task than studying communities directly.

There are many statistical approaches to learning the function $f$. The simplest approaches are linear regression methods that simply posit a model of the form $f(\vec{x}_i) = \beta_0 + \sum_{k=1}^p \beta_k x_{i,k}$, where the $\beta$ are regression coefficients. There are two major problems with this approach. First, if $n<<p$ we have many fewer data points than independent variables, which means that an ordinary least squares regression will almost certainly overfit and yield poor out of sample predictions (as determined by cross validation). The second related issue is that this approach gives equal weight to each entry (gene, taxon) in $\vec{x}$ and does not provide any dimensional reduction. One way to solve this problem is via regularization~\cite{hastie_elements_2016} where the model is optimized with an additional penalty term seeks to reduce the number of non-zero $\beta$ coefficients (see LASSO and Ridge Regression). Regularization provides a solution to the problem of selecting which regressors (entries of $\vec{x}_i$) provide the most predictive power while also avoiding overfitting. In situations where the levels of noise are not too high and a sparse solution (small number of non-zero $\beta_k$) do allow for good predictions, these regularization methods typically succeed~\cite{fraebel_evolution_2020}. However, if the noise levels are high or the underlying process is not sparse, then even these methods will fail. Care must be taken in diagnosing when such a regression works and when it does not, see~\cite{fraebel_evolution_2020,Gowda:2021ve}. 

A major shortcoming of the simple formulation outlined above it that it lacks any interaction terms (e.g., $x_{k,i} x_{l,i}$). Adding these terms to the regression above increases the number of independent variables from $p$ to $\sim p + p^2/2$. Given limited data $n$, it is typically not advised to take this approach. However, including such interaction terms is desireable given the utility of considering pairwise interactions in complex systems~\cite{bialek_rediscovering_2007,schneidman_weak_2006}. One way to proceed is to use linear regression approaches that use groups of independent variables as regressors. For example, principle component regression~\cite{hastie_elements_2016} uses principle components as independent variables in a linear regression. This is qualitatively similar to the unsupervised approach outlined above.

The models above are linear, and this aids in their interpretation. However, explicitly non-linear methods for estimating $f$ are also possible. When and why such approaches are more or less appropriate in the microbial context is not yet clear. However, decision tree based methods such as random forests have proven useful for relating taxonomic variation to host phenotypes in the microbiome~\cite{blanton_gut_2016}. Random forests can model complex non-linear relationships between regressors and response variables~\cite{hastie_elements_2016}, while retaining some interpretability by assigning `importance scores' to each independent variable. As a result these methods can be used to assess the impact of a given taxon or gene on the community function (in a statistical, not causal, sense). Random forest regressions fit many decision trees to bootstrapped replicates of the data and average the result. This averaging procedure, often called `bagging', reduces the variance of the prediction and as a result, high variance/low bias regression approaches can provide lower variance predictions. 

Finally, as the amount of high quality data on microbial communities increases, the applicability of more recently developed neural network supervised prediction methods~\cite{mehta_high-bias_2019} will become more appropriate. These methods typically contain many millions or even billions of parameters and therefore require reasonably sized data sets to train. Advances in methods like transfer learning~\cite{weiss_survey_2016}, where existing trained networks are trained to solve a new problem given some data, mean that the user need not start from scratch. The challenge will be what we can learn from these networks once they are used to approximate $f$. In many cases neural networks do not generalize well and are susceptible to small amounts of noise on the input variables. We regard the application and interpretation of these network approaches to communities a problem at the forefront. It may be that we need to reconsider how networks are trained to properly learn the salient features of the structure function problem in microbial communities~\cite{blazek_explainable_2021}.

\subsection*{What we do and do not learn from a statistical approach}

What can the ensemble approach coupled with a statistical analysis like the one described above teach us about communities? We contend that by looking at an ensemble of well-chosen communities, and learning the main statistical features of community structure that determine function we can begin to learn what general properties of communities must be present to admit their functional properties. A handful of recent studies have begun to show the power of this approach~\cite{Goldford:2018jf,blanton_gut_2016,raman_sparse_2019,Gowda:2021ve}. What we recover from these studies is what reproducible features of communities are retained across replicate consortia. These reproducible features can be regarded as `good variables' for predicting community function from structure. In some cases, understanding these variables lets us control or predict the functional properties of consortia in synthetic systems~\cite{Gowda:2021ve} and in hosts~\cite{blanton_gut_2016}. Ultimately, we hope this approach can be used to design, predict and control microbial consortia in engineered and wild contexts to address the existential threat of climate change.  

However, even when these statistical approaches succeed in predicting community structure from function we often do not understand why, at a mechanistic level, the prediction succeeds. For example, in the case of Gowda \emph{et al.}~\cite{Gowda:2021ve} the reason for the success of the regression from gene content to phenotypes is not entirely clear. Nor do we believe that the regression can predict the impact of gene gain or loss mutations. Similarly, in the case of Blanton \emph{et al.}~\cite{blanton_gut_2016} the precise metabolic role of each bacterial taxon in the stunting of the host is unclear. In essence, the statistical approach lets us find good variables for design and control of communities, but it does not, by itself, tell us why these are good variables. 

\section*{Future directions: evolutionary rules of the structure-function mapping}
\label{S:6}

So, while the ensemble approach can help us solve the structure-function problem the deeper question of why nature constructs communities the way it does remains. We argue that the answer to this question will require considering the eco-evolutionary basis of the observed structure function mapping. Addressing this question is subject enough for a separate manuscript. However, we hope that through the careful application of quantitative methods, like those discussed here, to some of the model systems discussed above, we can open the door to understanding how nature constructs dynamic functional consortia.

\bibliographystyle{model1-num-names}
\bibliography{references.bib}







\end{document}